\documentclass[10pt, twocolumn, twoside]{article}
\usepackage[utf8]{inputenc}
\usepackage[english]{babel}
\usepackage[T1]{fontenc}
\usepackage[noadjust]{cite}
\usepackage{babel}
\usepackage{amsmath,amssymb}
\usepackage{graphicx}
\usepackage{xcolor}
\usepackage{array}
\usepackage{tabularx}
\usepackage{csquotes}
\usepackage{blindtext}
\usepackage{hyperref}
\hypersetup{
    colorlinks=true, 
    linktoc=all,     
    linkcolor=black,  
    urlcolor=blue,
    citecolor=blue
}
\usepackage{subcaption}
\usepackage{indentfirst}
\usepackage[top=3cm, bottom=3cm, left=3cm, right=3cm]{geometry}
\usepackage{lineno}

\begin{document}

\twocolumn[{\centering{\LARGE Time-resolved purification of photon pairs from ultrasmall sources\par}\vspace{3ex}
	{Vitaliy Sultanov$^{1, 2*}$ and Maria V. Chekhova$^{1, 2}$\par}\vspace{2ex}
    {$^1$ Friedrich-Alexander Universität Erlangen-Nürnberg, Staudstrasse 7 B2, 91058, Erlangen, Germany\par}\vspace{2ex}
    {$^2$ Max-Planck Institute for the Science of Light, Staudtstrasse 2, 91058, Erlangen, Germany\par}\vspace{2ex}
    {$^*$ Corresponding author: vitaliy.sultanov@mpl.mpg.de\par}\vspace{2ex}
	\today\par\vspace{4ex}}
{\bfseries Generation of entangled photons through spontaneous parametric down-conversion (SPDC) from ultrasmall sources like thin films, metasurfaces, or nanoantennas, offers unprecedented freedom in quantum state engineering. However, as the source of SPDC gets smaller, the role of photoluminescence increases, which leads to the contamination of two-photon states with thermal background. Here we propose and implement a solution to this problem: by using pulsed SPDC and time distillation, we increase the purity and the heralding efficiency of the photon pairs. In the experiment, we increase the purity of two-photon states generated in a $7\,\mu$m film of lithium niobate from 0.002 to 0.99. With the higher purity, we were able to observe and characterize different polarization states of photon pairs generated simultaneously due to relaxed phase matching. In particular, we showed the presence of orthogonally polarized photons, potentially usable for the generation of polarization entanglement.\par}
\smallbreak
\medbreak
\par\vspace{2ex}]

Miniaturized sources of quantum photonic states are in the spotlight of quantum research as they are vital for the investigation of light-matter interaction at the nanoscale and the realization of quantum technologies with integrated photonic circuits. One of the leading trends is ``flat optics", involving ultrathin layers, down to a thickness of several atomic layers, and metasurfaces~\cite{Yu:2014}. In linear and nonlinear optics, flat optical devices already outperform their bulk counterparts~\cite{Krasnok2018}, especially in terms of tunability and multifunctionality~\cite{Ko2022, Chen2021}. `Flat' platforms are also promising sources of quantum light, including single-photon and two-photon states~\cite{Toth2019, Soln2021, Sharap2023}. Nanoscale sources of photon pairs mainly use spontaneous parametric down-conversion (SPDC) without momentum conservation~\cite{Okoth2019}, which gives unprecedented flexibility for the engineering of quantum entanglement in position-momentum~\cite{Okoth2020,Zhang2022}, time-frequency~\cite{Santiago-Cruz2021}, and polarization~\cite{Sultanov2022}, although at the cost of low generation efficiency. Researchers try out different materials and designs for nanoscale sources of quantum light~\cite{Guo2022}, investigating new approaches for generation rate enhancement, quantum state engineering, and adding multi-functionality~\cite{Marino2019, Santiago-Cruz2021_Nano, Jin2021, Duong2022, Santiago-Cruz2022, Saerens2023, Son2023}.

A huge advantage of nanoscale sources for producing high-dimensional entangled photons, apart from the freedom in the state engineering, is that such sources are free from most of the entanglement degradation mechanisms. For instance, due to the confined volume of nonlinear interaction, the dispersion effects are negligible. However, the signal-to-noise ratio is significantly reduced by the presence of background photoluminescence~\cite{Okoth2019, Marino2019}. Although highly-dimensional entangled photonic states are robust to noise to some extent~\cite{Zhu2021}, at the nanoscale the noise level is so high that it significantly lowers the purity of the generated two-photon state and makes it impossible to certify a high degree of entanglement.

Photoluminescence is an incoherent process, therefore its rate scales linearly with the thickness of the source and at nanoscale it is much brighter than SPDC, whose rate scales quadratically with the thickness. Typically, background thermal noise surpasses photon pair generation by several orders of magnitude. Because photoluminescence is isotropic and spectrally broadband, photon pairs can be filtered from it neither in space nor in polarization nor in frequency. Although photon pairs can still be observed via correlation measurements, such "noisy" sources of two-photon light are barely feasible for quantum applications requiring a high purity of the generated quantum light.

The two-photon state generated via nanoscale SPDC is a mixture of the pure highly entangled (multimode) two-photon state $|\Psi\rangle$ and a maximally mixed state of the photoluminescent background noise,
\begin{equation}
    \hat{\rho} = p\left|\Psi\right\rangle \left\langle\Psi\right| + \frac{1-p}{d^2}\mathbb{I}_{d^2},
    \label{White_noise_model}
\end{equation}
where $p$ is the probability of the pure state, $d$ the dimensionality, or the number of modes, and $\mathbb{I}_{d^2}$ the $d^2$-dimensional identity operator~\cite{Ecker2019}. The number of spectral modes is very large as photons occupy a broad spectral range. Under this condition, the purity of the mixed part is negligibly small~\cite{Nape2021}, and $p$ fully determines the purity of the generated state.

\begin{figure}[hbt!]
    \centering
          \includegraphics[width=0.8\linewidth]{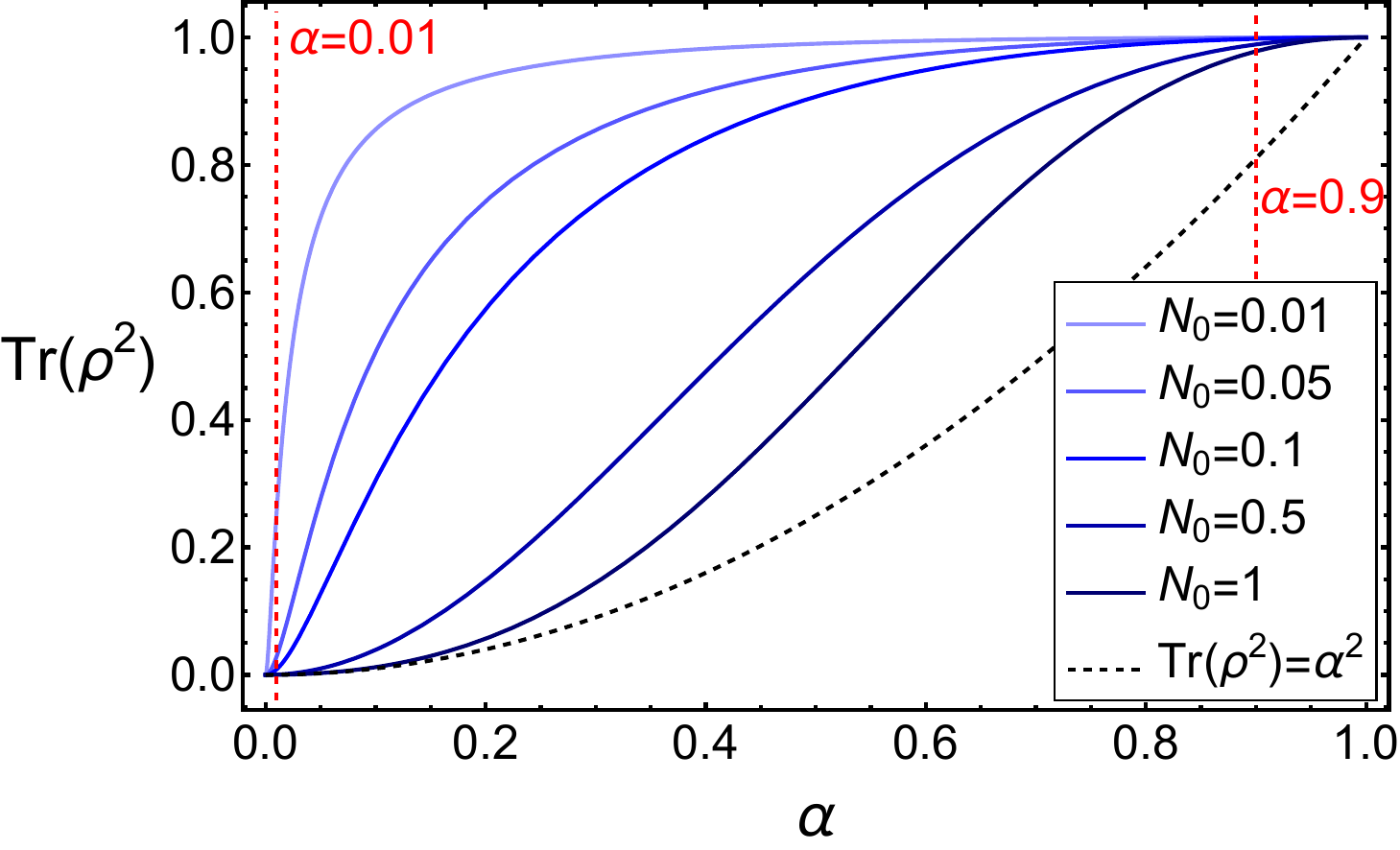}
    \caption{Purity of the two-photon state as a function of the fraction $\alpha$ of SPDC photons in the total number of photons $N_0$ for $d=1130$ modes. Red dashed lines show $\alpha$ obtained without ($\alpha=0.01$) and with ($\alpha=0.9$) the time-resolved distillation.}
    \label{Purity}
\end{figure}

For low-flux strongly multimode light, the probabilities to have a pair from SPDC and photoluminescence scale, respectively, linearly and quadratically with the corresponding mean photon numbers $N_{SPDC}, N_{PL}$:
\begin{equation}
    p =C N_{SPDC}, 1-p =C N_{PL}^2,
\end{equation}
where $C$ is the proportionality coefficient. Therefore, $p$ depends on the total mean number of photons $N_0$ and the fraction of photons produced by SPDC, $\alpha=N_{SPDC}/N_0$, as
\begin{equation}
    p(\alpha, N_0) = \frac{\alpha}{\alpha+(1-\alpha)^2N_0}.
    \label{p_dumb_formula}
\end{equation}
A rigorous calculation (Supplementary Information, section 3) yields a very similar result. The purity of state~(\ref{White_noise_model}),
\begin{equation}
    Tr\left(\rho^2\right)=p^2\left(1-\frac{1}{d^2}\right)+\frac{1}{d^2},
\end{equation}
becomes $p^2$ for a highly dimensional state, $d\gg 1$. Figure~\ref{Purity} shows the purity of the state as a function of $\alpha$ for different values of the total photon number $N_0$. In nanoscale SPDC experiments, typically $\alpha< 10^{-2}$ and the purity of photon pairs is very low.

The solution we propose relies on the fundamental difference between the two processes. While SPDC is a parametric process and occurs almost instantaneously, photoluminescence is a non-parametric process with the time dynamic defined by the matter relaxation. Here we show that the photon pairs can be distilled from the photoluminescent background by time-resolved detection under pulsed SPDC. Resolving the time dynamics of emission is not possible under continuous-wave (CW) pump excitation~\cite{Flagg2012}, which is typically used for nano-SPDC. To the best of our knowledge, this is the first work dedicated to pulsed SPDC at the nanoscale.

\begin{figure}[hbt!]
    \centering
    \includegraphics[width=1\linewidth]{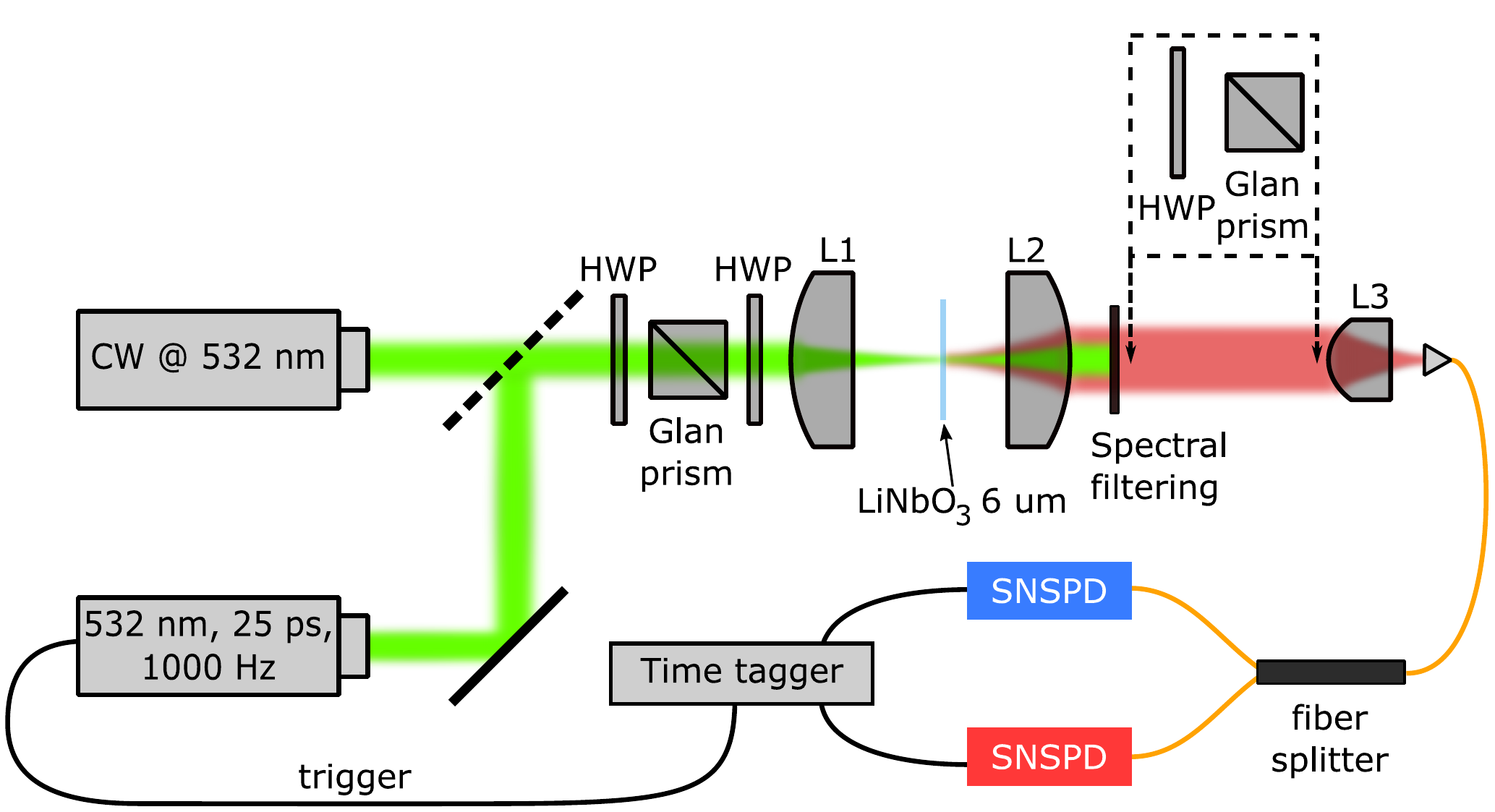}
    \caption{Laser radiation, either CW or pulsed, pumps a thin layer of lithium niobate to generate photon pairs. Two HWPs and a GP control the pump power and polarization. With a Hanbury Brown-Twiss setup (a fiber beam splitter, two single-photon detectors, and a time tagger), we register the generated light and analyze its two-photon correlations, heralding efficiency, and purity. Further on, another HWP and a GP  select different polarization states of the detected photons.}
    \label{Setup}
\end{figure}

As a source of photon pairs, we use a $7\,\mu m$ thick wafer of x-cut lithium niobate (LiNbO$_3$) illuminated by laser radiation with a wavelength of 532 nm, either CW or pulsed (Fig.~\ref{Setup}). For the experiments with the pulsed pump, we use a laser with 25 ps pulse duration and 1 kHz repetition rate. A set of two half-wave plates (HWP) and a Glan prism (GP) control the power and polarization of the pump. After focusing the pump onto the wafer, we collect the emitted photons, filter out the pump with a set of long-pass filters with a maximum cut-on wavelength of 950 nm and send photons to a Hanbury Brown - Twiss setup. The latter consists of a fiber beam splitter connected to two superconducting nanowire single-photon detectors (SNSPDs) and a time tagger, which registers the detectors' `clicks' and builds the distribution of the arrival time difference between the photon detections (`the coincidence histogram'). An additional set of a HWP and a GP filter an arbitrarily chosen linear polarization state of registered photons.

\begin{figure}[hbt!]
    \centering
    \begin{subfigure}{0.45\linewidth}
        \includegraphics[width=1\linewidth]{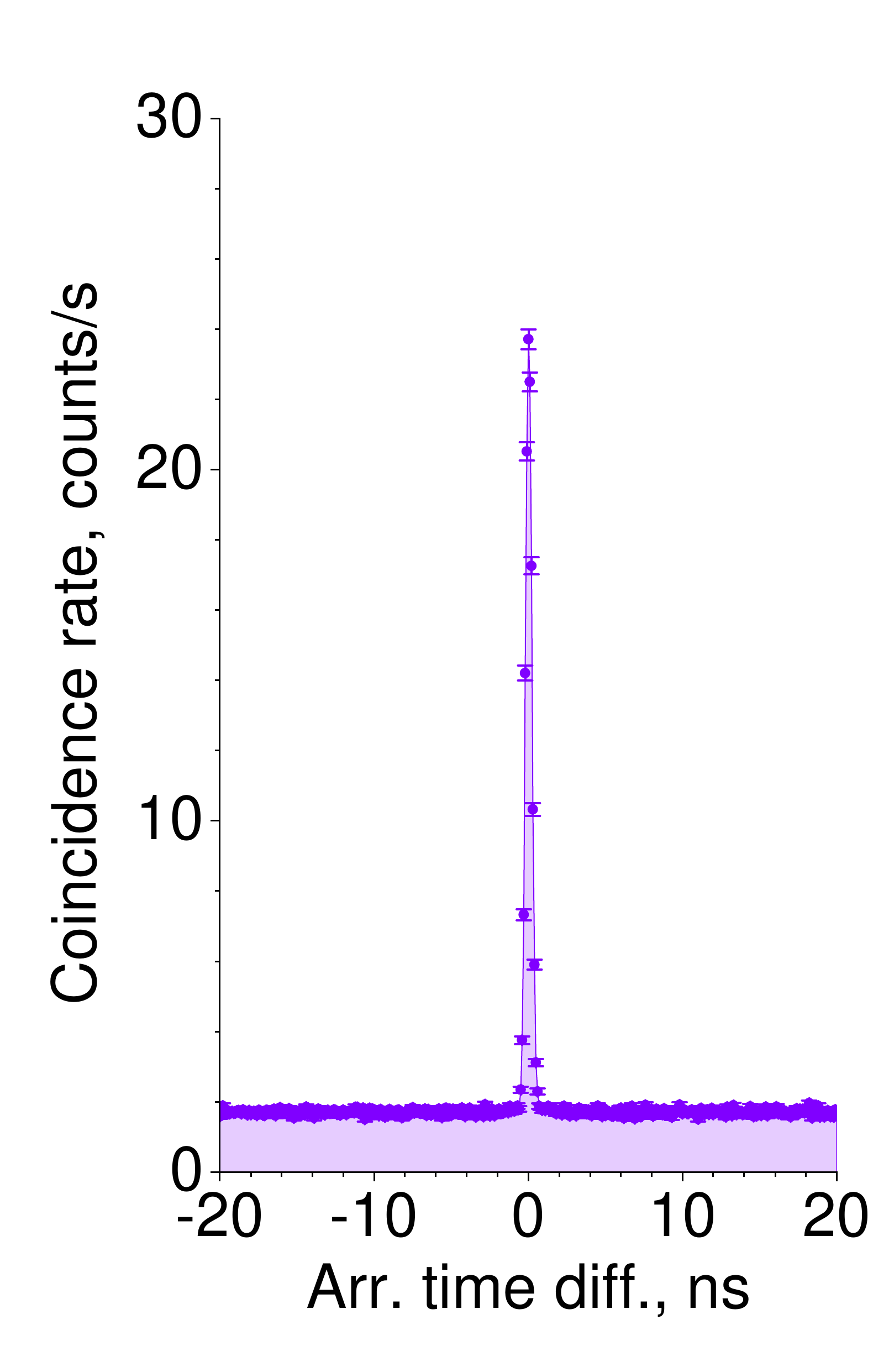}
        \caption{}
        \label{CW_example}
    \end{subfigure}
    \begin{subfigure}{0.43\linewidth}
        \includegraphics[width=1\linewidth]{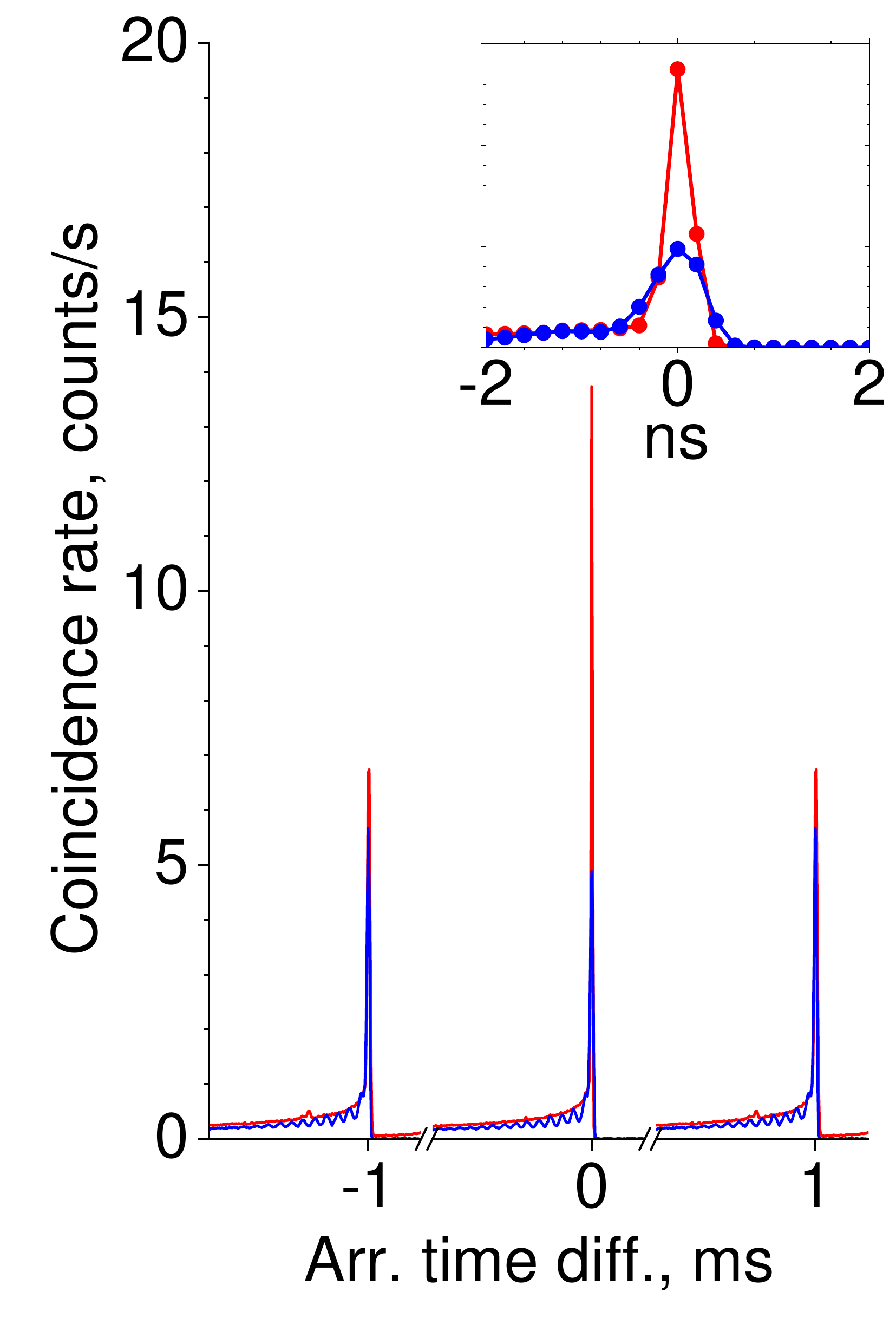}
        \caption{}
        \label{CSingles_example}
    \end{subfigure}
    \caption{In CW SPDC, the only timing information is in the distribution of the time delay between the arrivals of two photons, showing a sharp coincidence peak (panel \subref{CW_example}). In the pulsed regime, the distribution of the time delays between the counts of each detector (red and blue in panel \subref{CSingles_example}) and the pump pulse also shows sharp peaks, indicating SPDC photons. The long tails correspond to photoluminescence photons. The inset in panel \subref{CSingles_example} shows a zoom into the peak. By discarding the long tails in panel \subref{CSingles_example} we get rid of the contribution of photoluminescence and purify the detected two-photon state.}
    \label{Example}
\end{figure}

A typical coincidence histogram for the case of the CW pump is shown in Fig.~\ref{CW_example}. Although the pronounced narrow peak clearly indicates photon pair detection, there is a strong background of accidental coincidences, caused by the high rates of photons registered by both detectors. These rates, amounting to $1.2\cdot 10^5$ and $1.5\cdot 10^5$ s$^{-1}$, originate from photoluminescence and exceed the coincidence rate by several orders of magnitude. The ratio of the coincidence (after subtracting the accidentals) and singles rates is known as the {\it heralding efficiency}, 
and it is crucial for using SPDC as a source of single photons. In bulk SPDC sources, the heralding efficiency coincides with the detection efficiency. However, in the case of a noisy source, it also reflects the purity of the photon pairs. The heralding efficiency of each channel is related to $\alpha$ as $\eta_{1,2}=\alpha\;\eta_{1, 2}^{det}$, with $\eta_{1, 2}^{det}$ being the detection efficiency (see Supplementary Information, Section 3). Photoluminescence significantly lowers $\alpha$ and, as a result, the heralding efficiency.

To distill the photons emitted via SPDC from the thermal radiation caused by photoluminescence, we use a pulsed laser as a pump. Its electronic trigger synchronizes the detection of the emitted light, similar to the time-domain fluorescence lifetime imaging (FLIM) with single-photon counting~\cite{Becker2012}. Fig.~\ref{CSingles_example} shows the example of synchronous photon detection revealing the time dynamics of emission. We attribute high and sharp equidistant peaks to the emission of SPDC photons, whereas long subsequent ``tails" correspond to photoluminescence photons. By cutting the tails, we remove the contribution of photoluminescence to the single counts of both detectors and strongly suppress the rate of accidental coincidences. The coincidence histogram is obtained by acquiring the three-fold coincidences between the detectors' `clicks' and the electronic trigger of the laser (see Supplementary Information, section 2). 

\begin{figure}[hbt!]
    \centering
    \begin{subfigure}{0.8\linewidth}
        \includegraphics[width=1\linewidth]{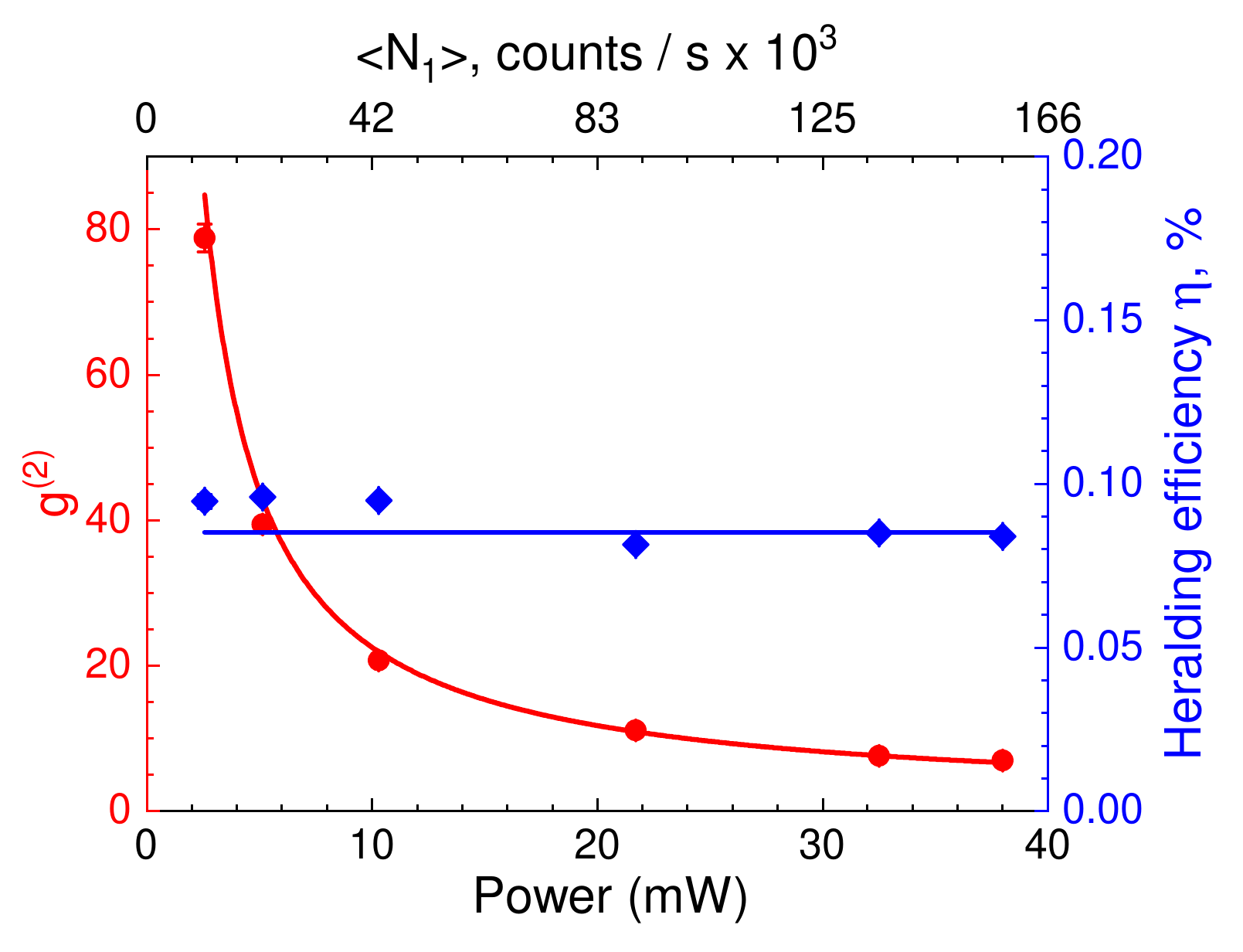}
        \caption{}
        \label{g2_HE_CW}
    \end{subfigure}
    \begin{subfigure}{0.8\linewidth}
        \includegraphics[width=1\linewidth]{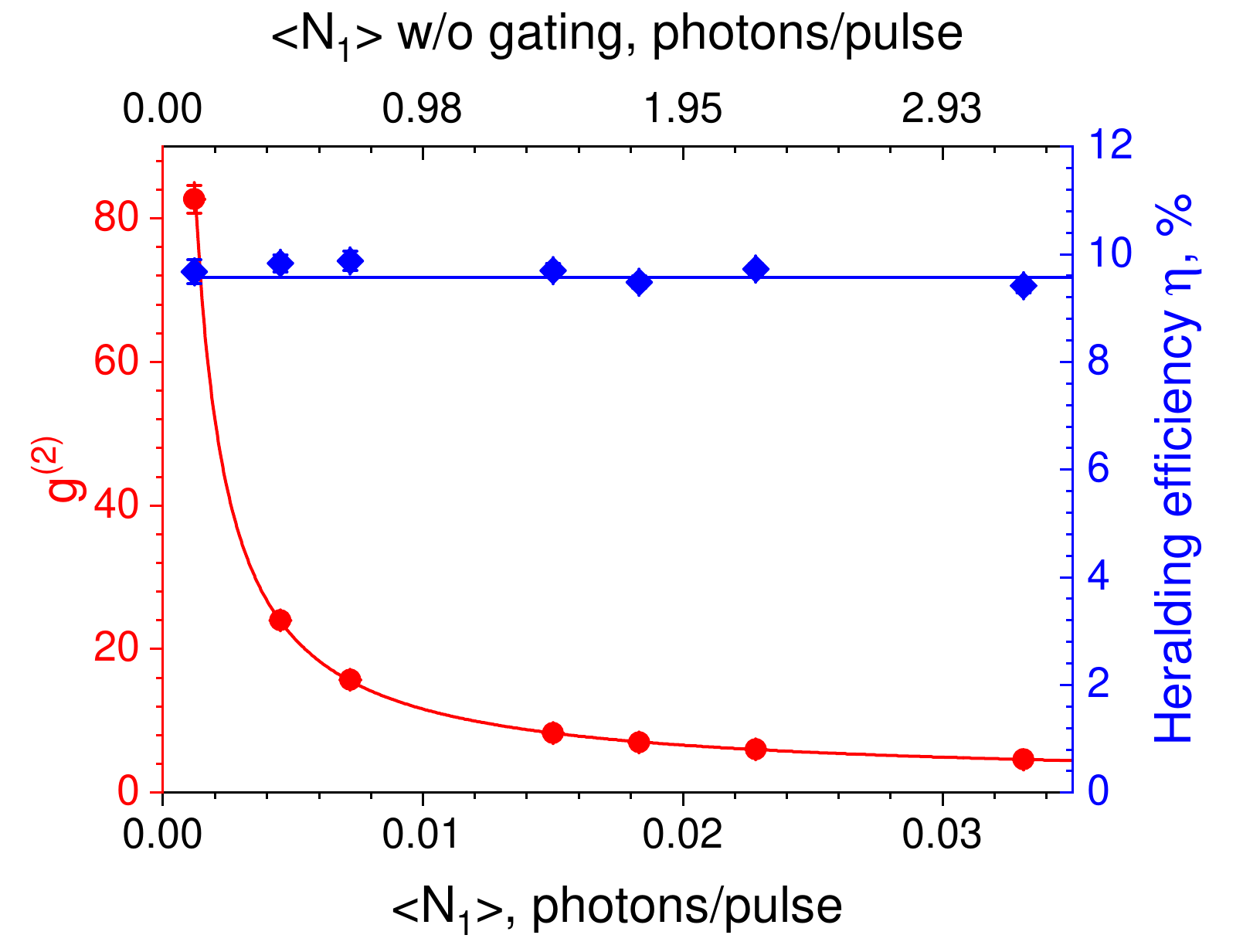}
        \caption{}
        \label{g2_HE_Pulsed}
    \end{subfigure}
    \caption{(\subref{g2_HE_CW}) Second-order normalized correlation function and heralding efficiency measured under CW pumping. Although $g^{(2)}$ is relatively high, the heralding efficiency is only a fraction of a percent, meaning that the probability of the photon pair detection if one detector clicks is extremely low. (\subref{g2_HE_Pulsed}) The same values measured under pulsed pumping and time-domain distillation. With roughly the same values of $g^{(2)}$, the heralding efficiency is much higher, now about 10\% instead of 0.1\%.}
    \label{g2_HE}
\end{figure}

To fairly compare photon pair generation in the CW and the pulsed regime, we acquire the statistics of single-photon and coincidence events for a set of input pump powers and calculate, for both cases, the second-order correlation function and the heralding efficiency (Fig.~\ref{g2_HE}). In both cases, we fit the experimental data with the inverse proportionality to the photon rate, which scales linearly with the pump power. Such dependence, as well as a high value of the second-order correlation function, clearly points towards photon pair detection. However, a high number of photons detected from the photoluminescent background in the CW case results in an extremely low heralding efficiency $0.085 \pm 0.002\%$. In contrast, in the pulsed regime, time-domain distillation increases the heralding efficiency to $9.6\pm0.1\%$, two orders of magnitude higher. We attribute the two orders of magnitude improvement in the heralding efficiency to the two-orders of magnitude higher value of $\alpha$ when the time-resolved distillation is applied. However, the absolute value of $\alpha$ is unknown yet. We determine it further from the correlation measurements with different polarization configurations of SPDC.
\begin{figure*}[htbp!]
    \centering
    \begin{subfigure}{0.24\linewidth}
        \includegraphics[width=1\linewidth]{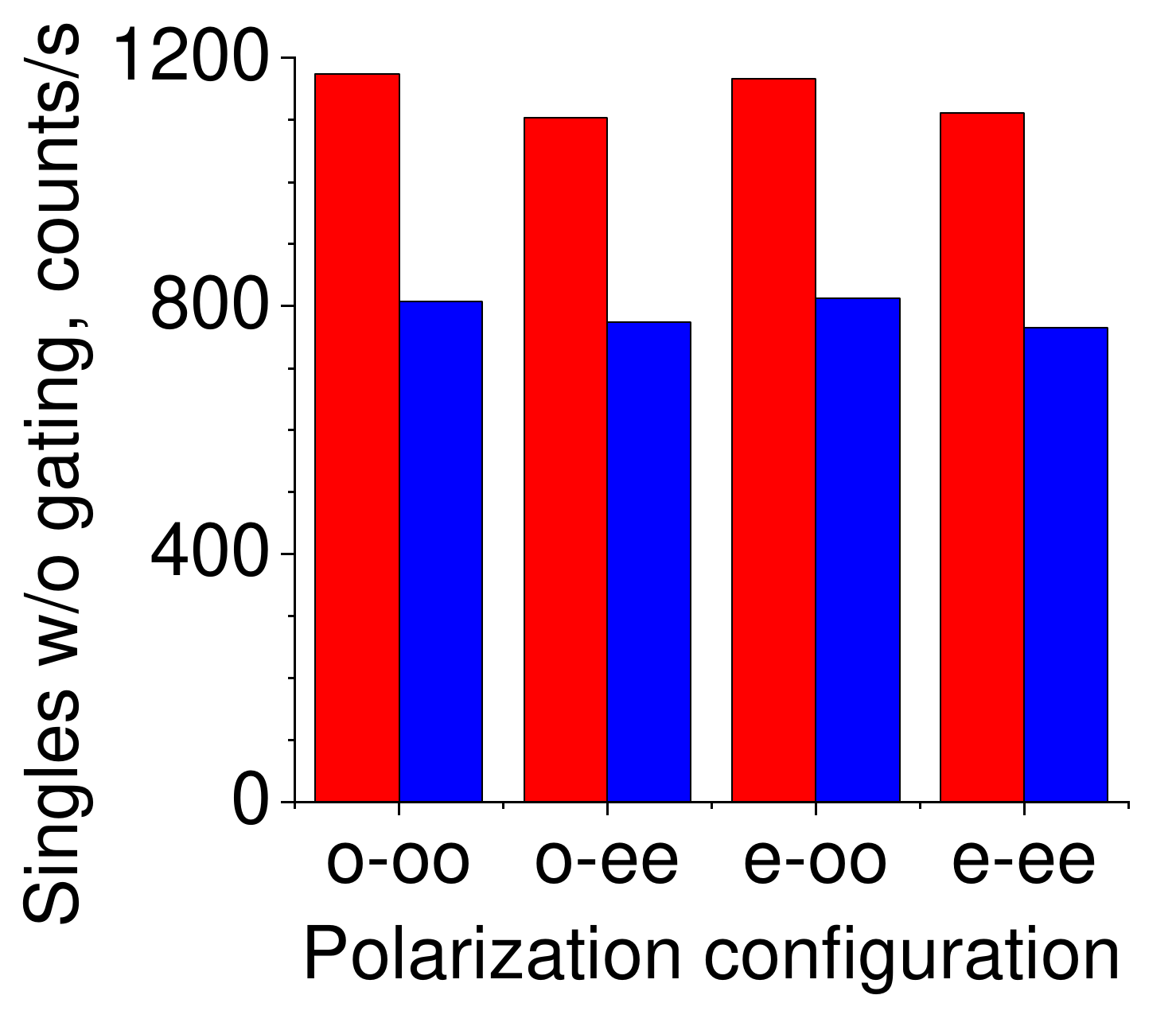}
        \caption{}
        \label{Singles_Histogram_polarization}
    \end{subfigure}
    \begin{subfigure}{0.24\linewidth}
        \includegraphics[width=1\linewidth]{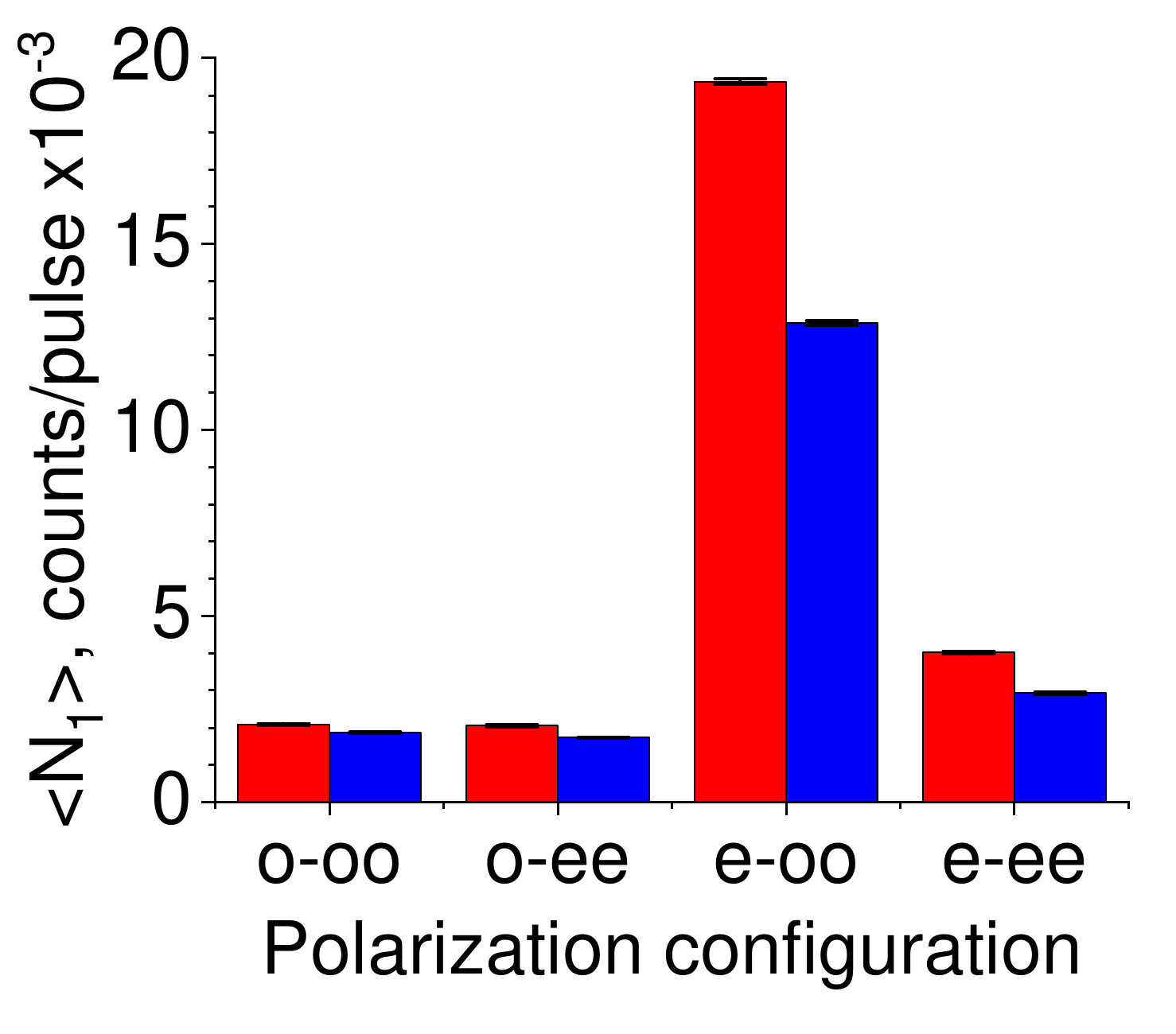}
        \caption{}
        \label{CSingles_Pulsed_polarization}
    \end{subfigure}
    \begin{subfigure}{0.24\linewidth}
        \includegraphics[width=1\linewidth]{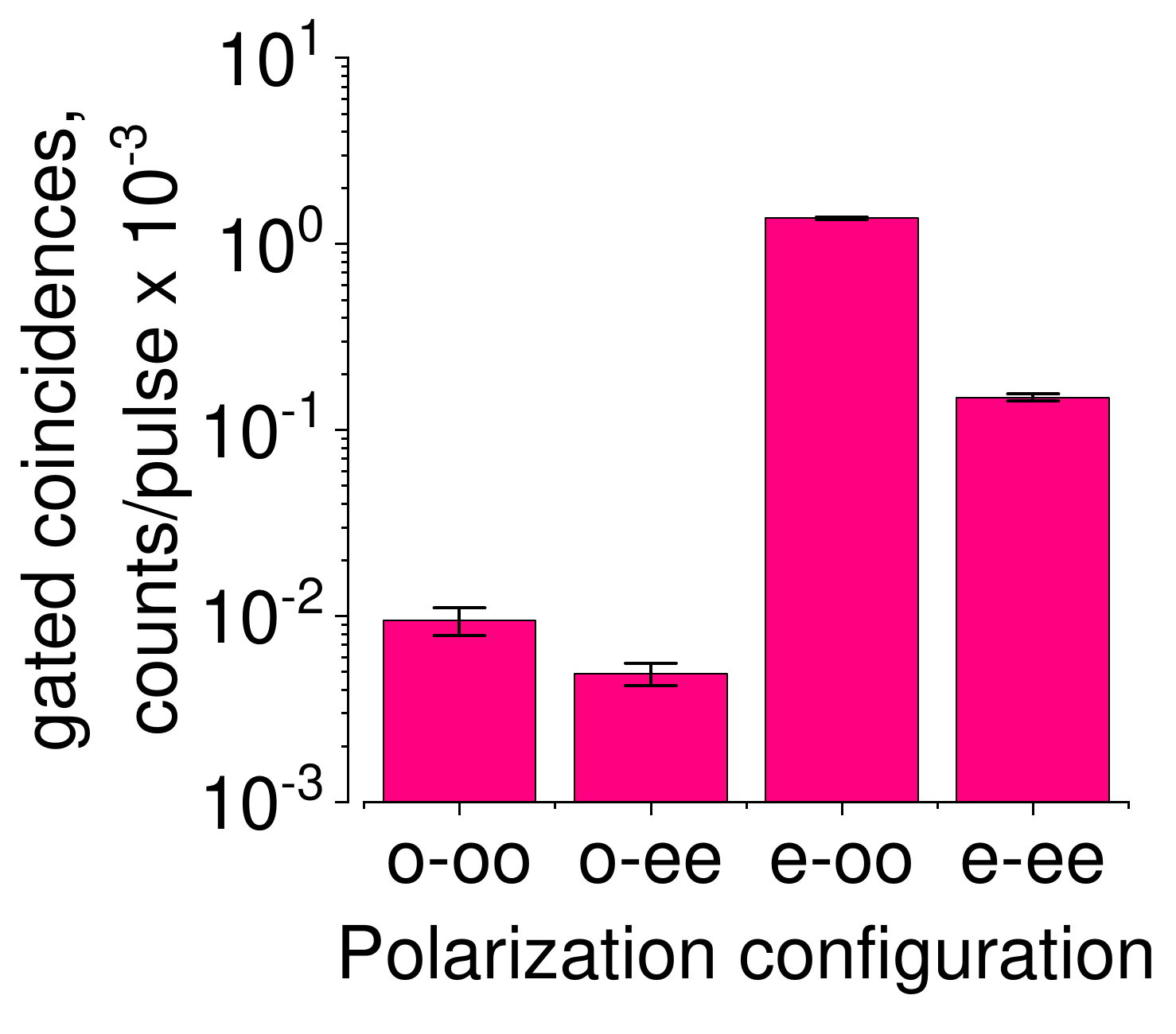}
        \caption{}
        \label{Coincidences_Pulsed_polarization}
    \end{subfigure}
    \begin{subfigure}{0.24\linewidth}
        \includegraphics[width=1\linewidth]{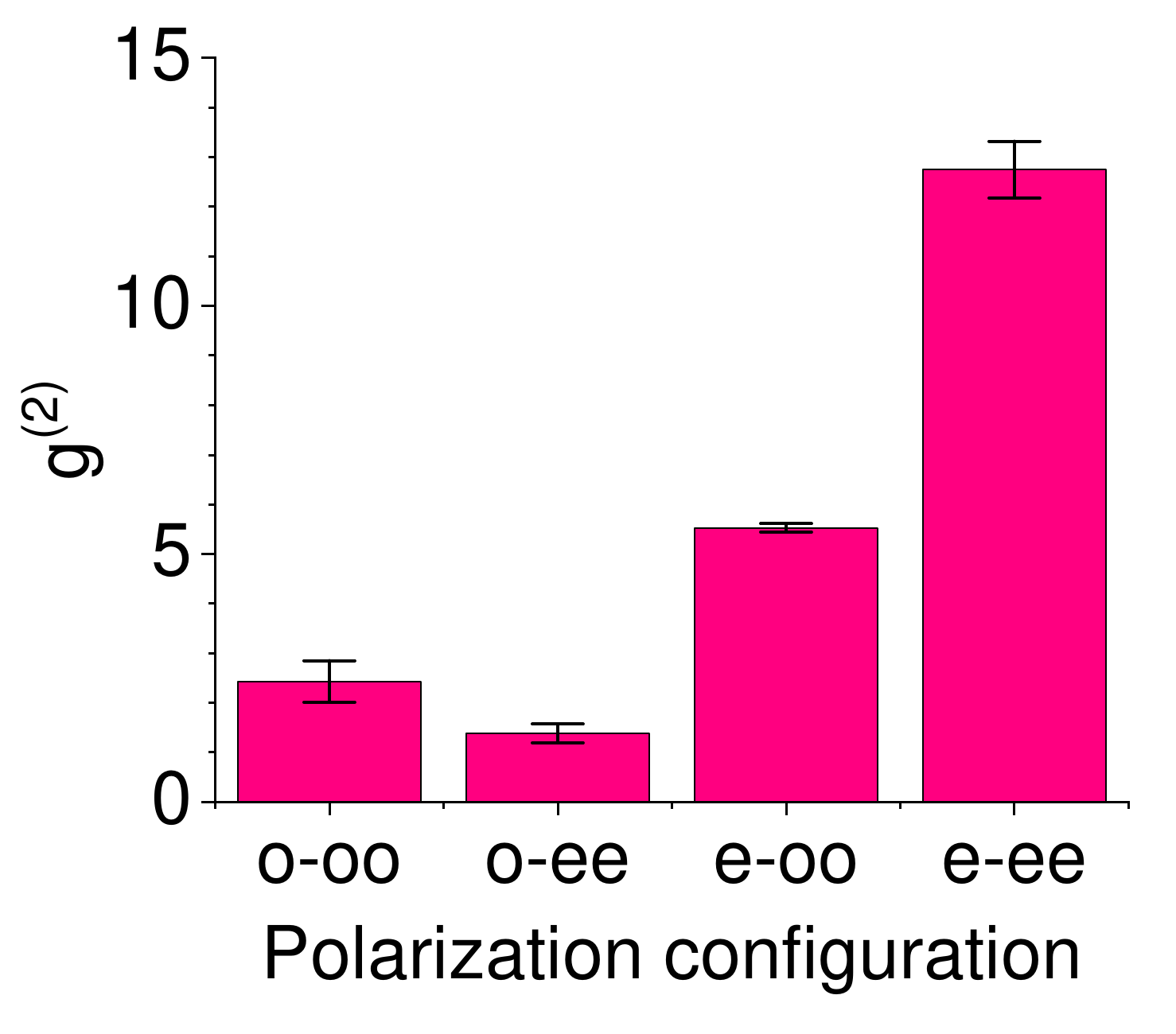}
        \caption{}
        \label{g2_Histogram_polarization}
    \end{subfigure}
    \caption{Rates of detected photons measured without (\subref{Singles_Histogram_polarization}) and with (\subref{CSingles_Pulsed_polarization}) time distillation for detector 1 (red) and 2 (blue); photon pair detection rates measured with time distillation (\subref{Coincidences_Pulsed_polarization}), and the resulting normalized second-order correlation function (\subref{g2_Histogram_polarization}) for different polarization configurations.}
    \label{Residual_PL}
\end{figure*}

Due to the relaxed phase matching condition in SPDC from ultrathin sources, pairs are generated both from ordinary (o-) and extraordinary (e-) polarized pump. The versatility of their polarization properties is only restricted by the efficiency of different types of SPDC, which can be adjusted by varying the source thickness, and the nonlinear tensor of LN, which has several nonzero elements (see Supplementary Information, section 5). We measure the rates of detected single photons and pairs (coincidences) for four different polarization configurations, involving e- and o-polarization of both the pump and detected photons. The results are shown in Fig.~\ref{Residual_PL} for both CW and pulsed SPDC. In the first case, there is no time distillation; therefore, the detected photons mainly come from photoluminescence and their rate does not depend on polarization (Fig.~\ref{Singles_Histogram_polarization}). In the second case, due to the time distillation, the rates of detected photons (Fig.~\ref{CSingles_Pulsed_polarization}) and coincidences (Fig.~\ref{Coincidences_Pulsed_polarization}) are strongly polarization-dependent. The corresponding values of $g^{(2)}(0,0)$ are shown in Fig.~\ref{g2_Histogram_polarization}. 

Because the coincidence rates contain almost no contribution from photoluminescence, we use them to analyze the polarization state of the pairs. The strongest coincidence count rate is from e-polarized pump. Despite the largest nonlinear tensor component $d_{33}$ supports the generation of e-polarized pairs (`e-ee' process), the rate of o-polarized pairs (`e-oo' process) is stronger because of the larger coherence length for this case (see Supplementary Information, section 5). Accordingly (see the inverse dependence of Fig.~\ref{g2_HE_Pulsed}), the second-order normalized correlation function $g^{(2)}$ is lower for o-polarized pairs than for e-polarized ones.

We notice an interesting feature: the co-existence of e-oo and e-ee processes leads to the coherent generation of $|oo\rangle$ and $|ee\rangle$ photon pairs, which, through two-photon interference, convert into pairs of orthogonally polarized photons~\cite{Kwiat1999}. This makes an ultrathin LN layer a promising source of polarization-entangled photons. Given its ultrabroad SPDC spectrum  (Supplementary Information, section 6), allowed by relaxed phase matching and inferring a high degree of time/frequency entanglement, such a source will provide high-dimensional hyper-entangled two-photon states.

Although o-polarized pump also generates pairs, the rates are 2 orders of magnitude lower (Fig.~\ref{Coincidences_Pulsed_polarization}). For o-ee, no SPDC pairs are expected since the effective value of $\chi^{(2)}$ is zero. The value of $g^{(2)}\approx 1$ for this configuration means that nearly all photons are produced by photoluminescence. For o-oo, $g^{(2)}$ is somewhat higher, indicating the presence of some SPDC photons.

By assuming the photon pair generation rate from the o-polarized pump to be negligible (see Supplementary Information, section 5), we estimate the residual photoluminescent background as the rate of single counts measured from this pump polarization. From Fig.~\ref{CSingles_Pulsed_polarization}, it is clear that the residual level of photoluminescent photons is only about 10\% of the overall single-count rate obtained after time distillation for the e-polarized pump. Therefore, the lower bound for $\alpha$ after time-domain distillation is 90\%, and a relatively low heralding efficiency of 10\% is entirely attributed to the detection efficiency. Then, based on the increase of the heralding efficiency after time-resolved distillation, we conclude that the fraction $\alpha$ of SPDC photons in the emitted light increases from 1\% to 90\%. This improvement leads to a significant increase of the purity of the generated two-photon state. Taking into account the detection efficiency of $10\%$, we conclude that the number of generated photons $N_0$ for the e-oo configuration is about 0.2 photons per pulse, and the number of modes $d$ can be estimated as 1130 (Supplementary Information, section 6). For this particular set of parameters, the purity increases from 0.002 to 0.99 (Fig.~\ref{Purity}). Therefore, time-resolved distillation allows us to increase the purity of the two-photon state to almost unity, making the emitted light feasible for quantum technology applications. 
 
In conclusion, we have implemented, for the first time, pulsed SPDC in an ultrathin source. We have shown that unlike the CW regime, pulsed regime enables time distillation of the detected photons and achieving their high purity. While the heralding efficiency for the CW case was only $0.085\pm 0.002\%$, in the pulsed regime, for the same value of the second-order correlation function, the measured heralding efficiency was two orders of magnitude higher, $9.6\pm 0.1\%$. This value is mainly limited by the detection inefficiency and optical losses (see Section 3 of the Supplementary Information). Through the time distillation, we increase the purity of the photon pairs from 0.01 to 0.99.

The estimated heralding efficiency is enough to use flat sources for real quantum technology applications such as quantum key distribution~\cite{Adachi2007} or boson sampling~\cite{Tillmann2013}. The small size of flat sources and relative freedom in the material choice makes non-phase-matched SPDC sources a convenient tool for the generation of quantum light under `flat' geometry. Due to the loose phase matching condition for nanoscale SPDC, any nonlinear materials can be used for photon pair generation. It is of particular interest to test monolayers~\cite{MoS_monolayer} and few-layer crystals~\cite{Guo2022}. Such materials possess an extremely high value of second-order nonlinearity and maintain all the unique features of nanoscale SPDC in the extreme case of vanishing crystal length. Further, one can create composite materials and combine ultrathin sources of photon pairs with, i.e., quantum dots, to perform various quantum operations. This requires a high heralding efficiency of the two-photon source, which is available from flat sources in the pulsed regime.

\end{document}


\begin{centering}
    \LARGE Time-resolved purification of photon pairs from ultrasmall sources: Supplementary Information\par\vspace{3ex}
\end{centering}
\begin{centering}
    {Vitaliy Sultanov$^{1, 2*}$ and Maria V. Chekhova$^{1, 2}$\par}\vspace{2ex}
    {$^1$ Friedrich-Alexander Universität Erlangen-Nürnberg, Staudstrasse 7 B2, 91058, Erlangen, Germany\par}\vspace{2ex}
    {$^2$ Max-Planck Institute for the Science of Light, Staudtstrasse 2, 91058, Erlangen, Germany\par}\vspace{2ex}
    {$^*$ Corresponding author: vitaliy.sultanov@mpl.mpg.de\par}\vspace{2ex}
	\today\par\vspace{4ex}
\end{centering}

\section{Measurement of the second-order correlation function and the heralding efficiency for CW and pulsed SPDC}

In the CW case, the value of the stationary normalized second-order correlation function at the zero delay $g^{(2)}(\tau=0)$ can be found as \cite{Mandel_Wolf}
\begin{equation}
    g^{(2)}(\tau=0) =\frac{R_c}{R_1R_2T},
    \label{g_CW}
\end{equation}
where $R_c$ is the rate of photon coincidences measured within the coincidence window $T$, and $R_1$ and $R_2$ are the rates of single counts from two detectors. 

The heralding efficiency can be found as
\begin{equation}
    \eta_{1,2} = \frac{R_c-R_{acc}}{R_{2,1}},
   \label{eta_CW}
\end{equation}
where $R_{acc} = R_1R_2T$ is the rate of accidental coincidences within the coincidence time window. The index 1 or 2 refers to the heralding probability in the first or second channel, respectively. 

In the pulsed regime, it is convenient to operate with numbers of single counts and coincidences per pulse, which are found from the synchronous detection of the single counts and coincidences. The second-order correlation function $g^{(2)}(t_1=0, t_2=0)$ is then found as~\cite{Ivanova2006}
\begin{equation}
    g^{(2)}(t_1=0, t_2=0) = \frac{\langle N_c\rangle}{\langle N_1\rangle\langle N_2\rangle},
       \label{g_pulsed}
\end{equation}
with $\langle N_c\rangle$, $\langle N_1\rangle$, and $\langle N_2\rangle$ being the average numbers of coincidences and single counts per pulse. We find the number of single counts by integrating over the peak at the zero delay (Fig.~\ref{CSingles_example}), eliminating the tail of photoluminescent photons. Thus, we are able to eliminate the vast majority of the photoluminescent background and distill photon pairs. The width of the peak is determined by the detection jitter and also defines the effective coincidence window. We then use the same time window to find the number of three-fold coincidences $\langle N_c\rangle$. The heralding efficiency is then found as
\begin{equation}
    \eta_{1,2} = \frac{\langle N_c\rangle - \langle N_{acc}\rangle}{\langle N_{2,1}\rangle},
       \label{eta_pulsed}
\end{equation}
with the average number of accidental coincidences per pulse $\langle N_{acc}\rangle=\langle N_{1}\rangle\langle N_{2}\rangle$.

\section{Synchronous coincidence detection in the pulsed regime}
To distill photon pairs from photoluminescence, we perform the synchronous detection of photon pairs generated via pulsed SPDC. We use the electronic trigger from the pump laser to synchronize the detection of photons with the pump pulses. Whereas in the CW regime, the only time information available is the arrival time difference between the clicks of two detectors, synchronous detection allows us to relate the detection time with the time of the pump pulses interacting with a crystal.
\begin{figure}[hbt!]
    \centering
    \begin{subfigure}{0.37\linewidth}
        \includegraphics[width=1\linewidth, valign=b]{CSingles_example.pdf}
        \caption{}
        \label{CSingles_example}
    \end{subfigure}
    \hspace{1cm}
    \begin{subfigure}{0.5\linewidth}
        \includegraphics[width=1\linewidth, valign=b]{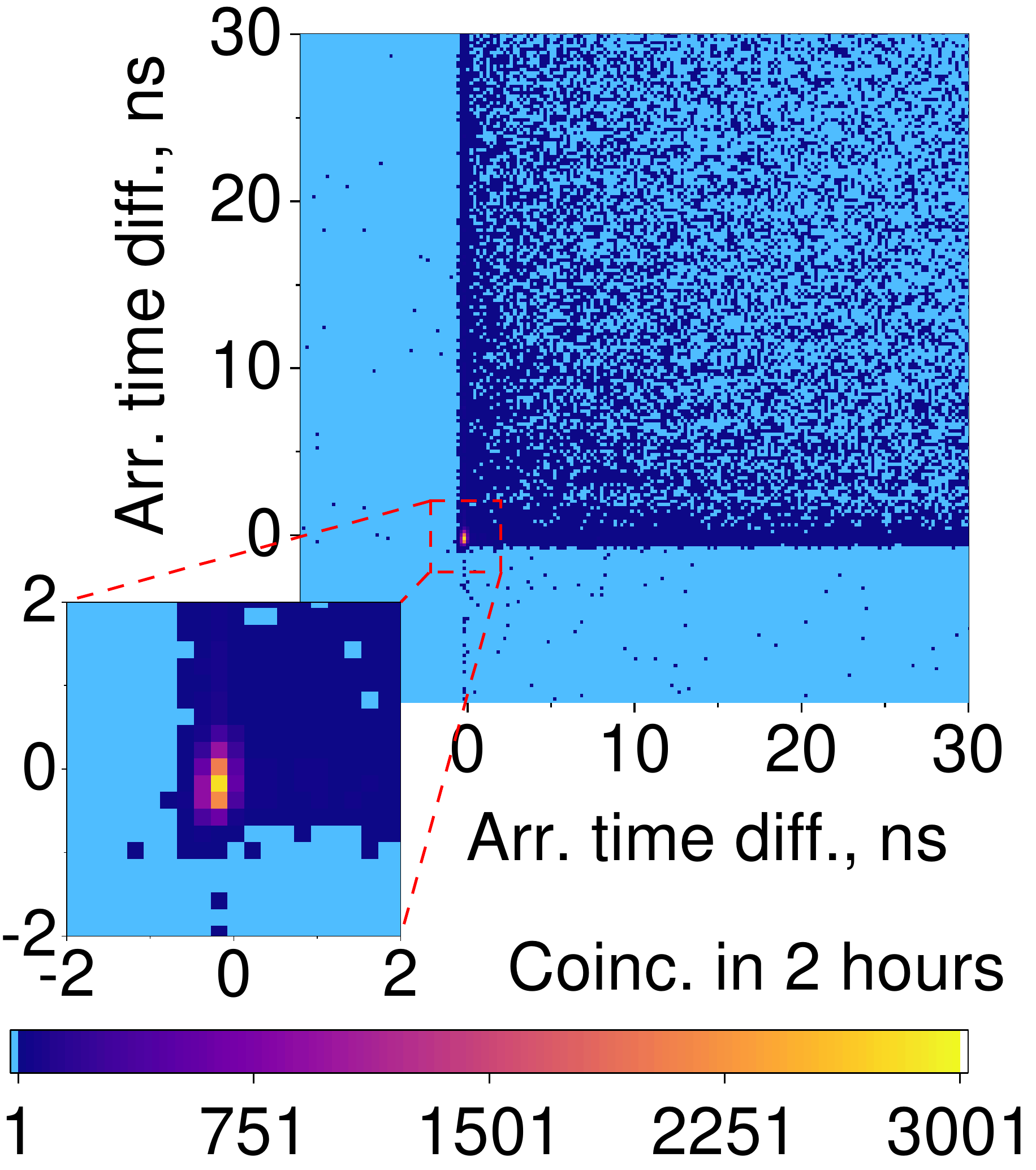}
        \caption{}
        \label{2DHistogram_example}
    \end{subfigure}
    \caption{The histogram of coincidences between the single clicks of one of the detectors and teh laser trigger (Fig. \subref{CSingles_example}) reveals the time dynamics of the emission. High sharp equidistant peaks corrspond to teh emission of photon pairs generated via SPDC, whereas subsequent tails originate from photoluminescence. The correlation measurements are performed with the three-fold coincidences between the clicks of both detectors and the laser trigger (Fig. \subref{2DHistogram_example})}
    \label{Example}
\end{figure}
We acquire the histogram of coincidences between the single click of one of the detectors and the laser trigger (Fig.~\ref{CSingles_example}), similar to the time-domain fluorescence lifetime imaging (FLIM) with single-photon counting~\cite{Becker2012}. We attribute the sharp equidistant peaks to photons generated via SPDC, while the following "tails" are caused by photoluminescence. By eliminating the photoluminescent tails, we register only single-photon counts related to SPDC. This procedure is similar to gating the detectors, with the gating time equal to the width of the peaks. For the correlation measurements, we register three-fold coincidences between both detectors and the electronic trigger. As a result, we acquire a two-dimensional histogram of coincidences (Fig.~\ref{2DHistogram_example}). The histogram features a high and narrow peak of coincidences at near-zero delay, corresponding to the detection of photon pairs. We select the area of the histogram corresponding to the overlap of the central peaks in Fig.~\ref{CSingles_example} and consider the coincidences in this area as real coincidences. Therefore, the procedure involves the time filtering of photon counts and allows us to perform time-resolved distillation of photon pairs. 

To calculate the heralding efficiency and the second-order correlation function in the CW and pulsed regime, we measure the number of single counts and coincidences per pump pulse and use equations (\ref{g_CW},\ref{eta_CW}) and (\ref{g_pulsed},\ref{eta_pulsed}), respectively. It is worth mentioning that the efficiency of time-resolved purification strongly depends on the time resolution of the detectors. The obtained results can be improved even further by using detectors with a smaller jitter.

\section{The relation between purity and heralding efficiency}

Here we show that a rigorous calculation of the two-photon state purity, taking into account the multi-mode nature and photon-number statistics of SPDC and photoluminescence, results in an expression similar to one obtained with the simple model in the main text. To find the statistical properties of the generated light, we use the generating function $Q(x) = \sum_nP(n)(1+x)^n$, where $P(n)$ is the probability of having $n$ photons~\cite{Klyshko}. The $n$-th order correlation function and the probability $P(n)$ can be calculated, respectively, as
\begin{gather}
    G^{(n)}(\tau=0) = \frac{\partial^n Q(x)}{\partial x^n}|_{x=0},\\
    P(n) = \frac{1}{n!}\frac{\partial^n Q(x)}{\partial x^n}|_{x=-1}.
\end{gather}
Assuming that we have the mixture of light generated via SPDC and photoluminescence, we are interested in the probability $p$ of finding a photon pair generated via SPDC exclusively, which determines the purity of the state. Let us find its relation to the fraction $\alpha$ of SPDC photons in the mixture, which can be determined by measuring correlations up to $G^{(2)}$ in this case~\cite{Goldschmidt2013}. We start from the derivation of the generating function for SPDC and photoluminescence. The states generated via SPDC in two conjugated modes $1,2$ and via photoluminescence in one mode are, respectively~\cite{Zhu2021}, 
\begin{gather}
    \left|\Psi\right\rangle_{SPDC}=\sum^{\infty}_{m=0}\left(\frac{\mu_{SPDC}}{1+\mu_{SPDC}}\right)^{\frac{m}{2}}\frac{a_{1}^{\dagger m}a_{2}^{\dagger m}}{m!(1+\mu_{SPDC})^{1/2}}\left|vac\right\rangle,\\
    \hat{\rho}_{PL} = \sum_{n=0}^{\infty}\left(\frac{\mu_{PL}}{1+\mu_{PL}}\right)^{n}\frac{1}{1+\mu_{PL}}\left|n\right\rangle\left\langle n\right|.
\end{gather}
Here, $\mu_{SPDC}$ and $\mu_{PL}$ are the mean photon numbers in SPDC and photoluminescence modes, respectively. The probabilities to have $n$ and $m$ photons in the selected modes are
\begin{gather}
    P_{SPDC}(n, m) = \delta_{nm}\frac{\mu_{SPDC}^n}{\left(1+\mu_{SPDC}\right)^{n+1}},\\
    P_{PL}(n) = \frac{\mu_{PL}^n}{\left(1+\mu_{PL}\right)^{n+1}},
\end{gather}
where $\delta_{nm}$ is the Kronecker delta. These probabilities allow us to construct the corresponding generating function for two-mode SPDC light and single-mode photoluminescence. Further on, as the modes of photoluminescence are independent, the generating function of photoluminescence emitted in the same two modes as SPDC is the product of single-mode generating functions for each of these modes~\cite{Klyshko}. Then, the two-mode generating functions for SPDC and photoluminescence are
\begin{gather}
    Q_{SPDC}(x, y) = \sum_{n,m=0}^{\infty}P_{SPDC}(n, m)\left(1+x\right)^{n}\left(1+y\right)^m = \frac{1}{1-\mu_{SPDC}\left(x+y+xy\right)},\\
    Q_{PL}(x, y) = \sum_{n,m=0}^{\infty}P_{PL}(n)\left(1+x\right)^n P_{PL}(m)\left(1+y\right)^m = \frac{1}{\left(1-\mu_{PL}x\right)\left(1-\mu_{PL}y\right)}.
\end{gather}
Since photoluminescence and SPDC are two independent processes, the resulting generating function for the selected pair of modes $1,2$ is also the product of the generating functions of SPDC and photoluminescence:
\begin{equation}
    Q_{1,2}(x, y) = Q_{SPDC}(x, y)Q_{PL}(x, y).
\end{equation}
Finally, we take into account the multi-mode nature of photoluminescence and SPDC. For the simplicity of calculations, we assume that all modes are populated equally with $\mu_{SPDC}$ photons per mode on average for SPDC and $\mu_{PL}$ photons per mode for photoluminescence. This assumption is realistic in the case where the spectra of SPDC and photoluminescence are extremely broad and flat~\cite{Okoth2019}. Then, assuming that SPDC is emitted in $d$ pairwise correlated modes, and photoluminescence is equally distributed over the same $2d$ modes, the resulting generating function for the emitted light is
\begin{equation}
    Q(x, y) = \left(Q_{1,2}(x, y)\right)^d = Q^d_{SPDC}(x, y)Q_{PL}^d(x, y).
    \label{Q_final}
\end{equation}

From generating function~(\ref{Q_final}), we find the probability of a photon pair as
\begin{multline}
    P(1, 1) = \frac{\partial^2 Q(x, y)}{\partial x\partial y}|_{x=0,\, y=0}  \\
    =\frac{1}{\left(1+\mu_{SPDC}\right)^d}\frac{1}{\left(1+\mu_{PL}\right)^{2d}}\left[\frac{d\mu_{SPDC}}{1+\mu_{SPDC}}+\left(\frac{d\mu_{PL}}{1+\mu_{PL}}\right)^2\right].
\end{multline}
The same expression can be obtained by calculating this probability as
\begin{equation}
 P(1, 1) = P_{SPDC}(1,1)P_{PL}(0,0)+P_{SPDC}(0,0)P_{PL}(1,1).
 \label{prob}
\end{equation}
The purity of the generated two-photon state can be then found as the probability that the photon pair comes from SPDC,
\begin{equation}
    p = \frac{P_{SPDC}(1,1)P_{PL}(0,0)}{P(1, 1)} = \frac{\frac{d\mu_{SPDC}}{1+\mu_{SPDC}}}{\frac{d\mu_{SPDC}}{1+\mu_{SPDC}}+\left(\frac{d\mu_{PL}}{1+\mu_{PL}}\right)^2}.
\end{equation}
We can now write this expression in terms of the total number of photons in $d$ modes, $N_0 = d(\mu_{SPDC}+\mu_{PL})$, and the fraction $\alpha$ of SPDC photons in the total number of photons:
\begin{equation}
    p\left(N_0,\, \alpha,\, d\right) = \frac{\alpha}{\alpha +(1-\alpha)^2 N_0\kappa},
    \label{p_MM_formula}
\end{equation}
where 
\begin{equation}
    \kappa=\frac{1+\alpha N_0/d}{(1+(1-\alpha) N_0/d)^2}.
\end{equation}
In the limit of small numbers of photons per mode, where $N_0/d=\mu_{SPDC}+\mu_{PL}<<1$, $\kappa\rightarrow1$ and Eq.~(\ref{p_MM_formula}) approaches the result (4) of the simple model described in the main manuscript.

\begin{figure}[hbt!]
    \centering
    \begin{subfigure}{0.45\linewidth}
        \includegraphics[width=1\linewidth, valign=b]{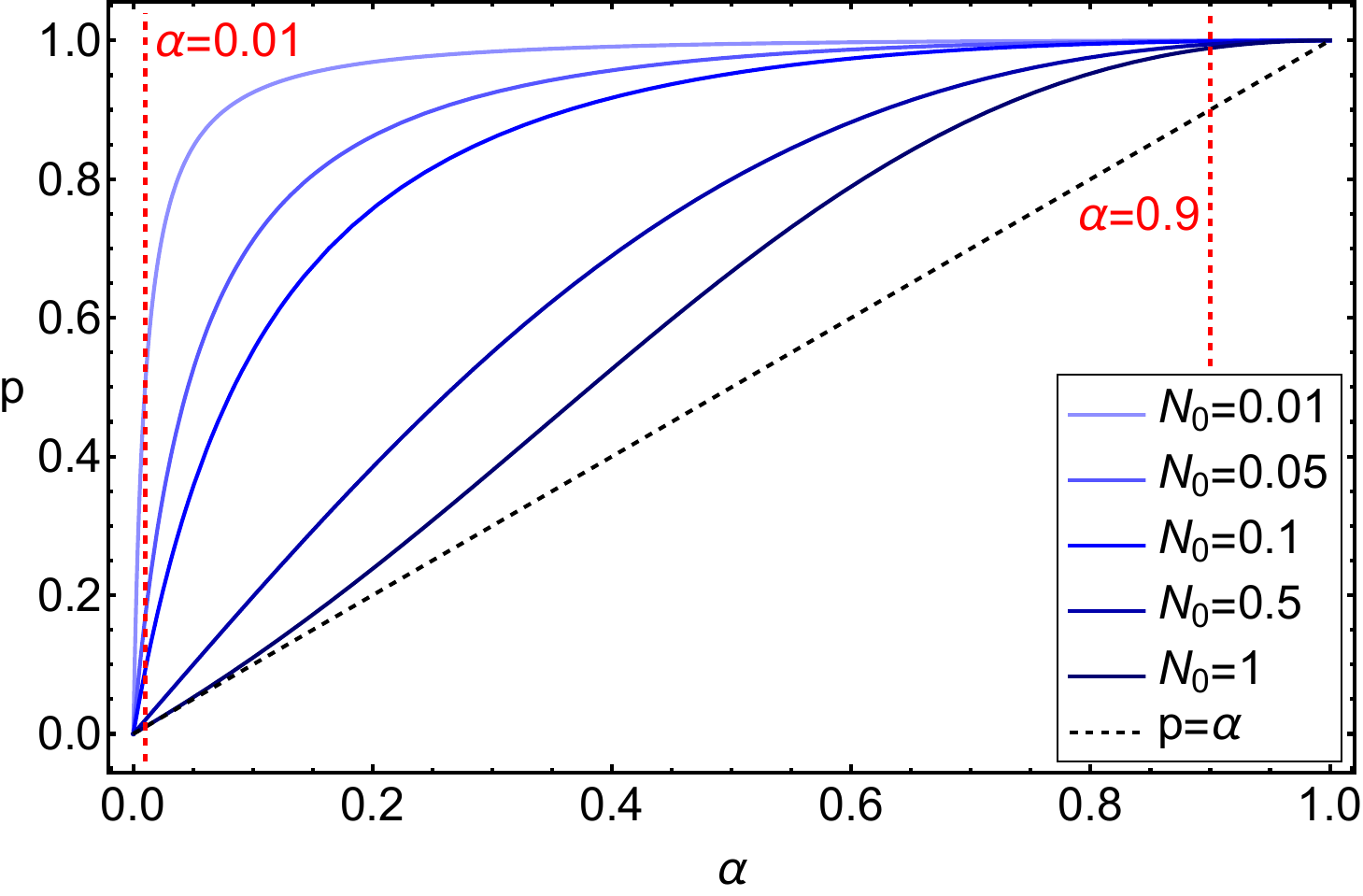}
        \caption{}
        \label{P_dumb}
    \end{subfigure}
    \hspace{1cm}
    \begin{subfigure}{0.45\linewidth}
        \includegraphics[width=1\linewidth, valign=b]{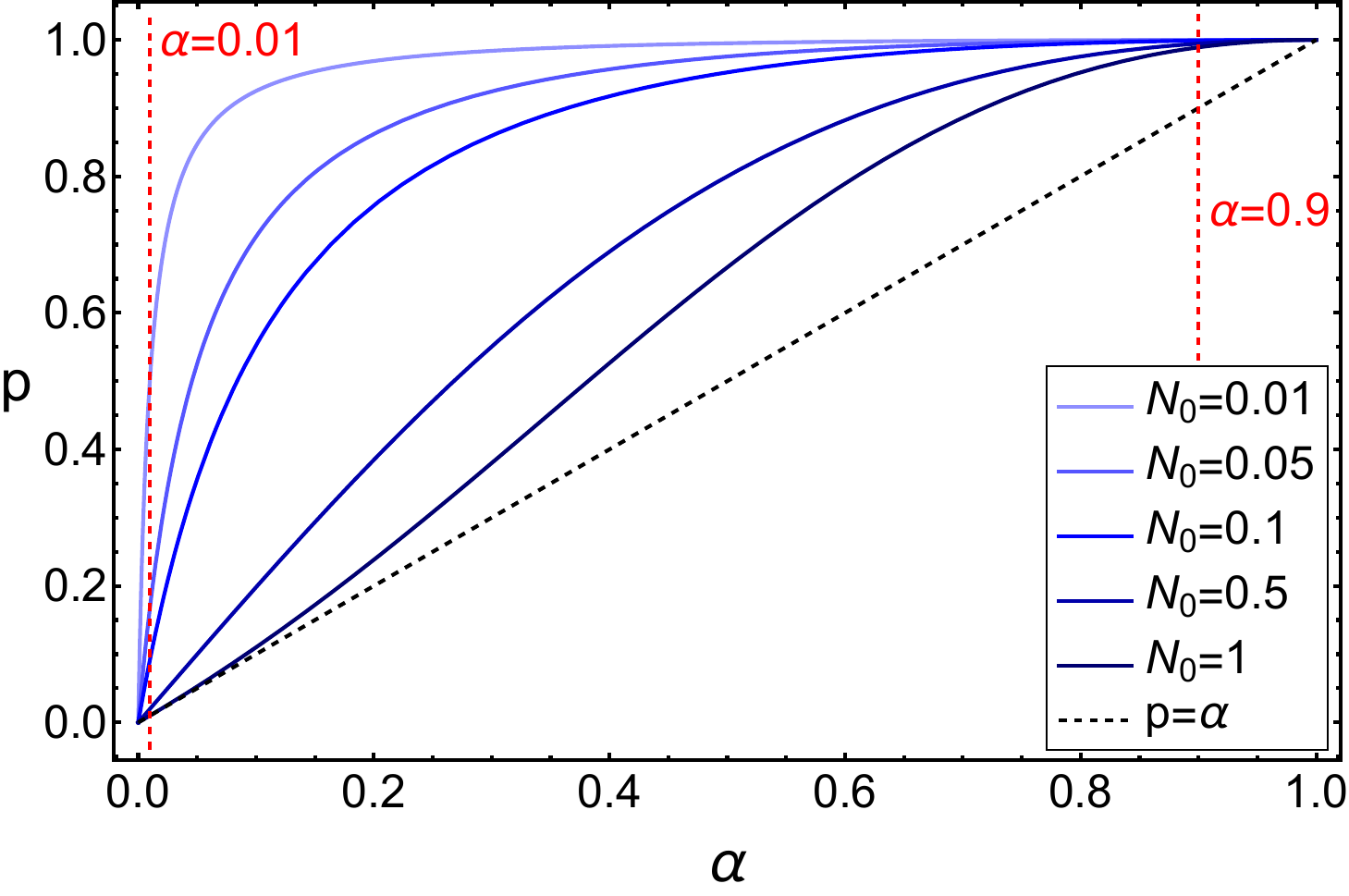}
        \caption{}
        \label{P_MM}
    \end{subfigure}
    \caption{For both the simple model (\subref{P_dumb}) and the multimode consideration (\subref{P_MM}) with $d=1130$, the two-photon state purity $p$ as a function of $\alpha$ and $\mu_0$ behaves in a similar way. The red dashed lines show the value of $\alpha$ without ($\alpha=0.01$) and with ($\alpha=0.9$) the time distillation applied.}
    \label{P}
\end{figure}

In Fig.\ref{P}, we show $p$ as the function of $\alpha$ obtained with the simple model (Fig.\ref{P_dumb}) and by accounting for the multimode nature of radiation (Fig.\ref{P_MM}). For the second case, we estimate $d=1130$ based on the spectral measurements described below in Section 5. Moreover, we consider the case of $N_0<1$, where the probability of more than two photons is relatively low. Under this limitation, $p$ is always considerably higher than $\alpha$, and so is the purity of the generated state (Fig. 1 in the main manuscript). It must be said that we consider the purity of the generated state taking the losses of the detection system out of the scope. This allows us to neglect the probability of having a pair consisting of one photon generated via SPDC and the other photon generated via photoluminescence. If the losses are taken into account, these probabilities must be also considered.

There are three parameters that determine the probability of the SPDC-generated photon pair and the purity of the generated two-photon state: the total number of photons $N_0$, the fraction of SPDC photons $\alpha$, and the number of modes $d$. The first two parameters can be determined in the experiment by measuring the numbers of single counts $\langle N_{1,2}\rangle$ and coincidences $ G^{(2)}_{12} = \langle:N_{1}N_{2}:\rangle$ in the HBT experiment. From the generating function (\ref{Q_final}) we get
\begin{gather}
    N_0 = \langle N_{1}\rangle = \langle N_{2}\rangle = \frac{\partial Q(x, y)}{\partial x}|_{x=0,\,y=0}=d(\mu_{SPDC}+\mu_{PL}),\\
    G^{(2)}_{12} \equiv \frac{\partial^2 Q(x, y)}{\partial x\partial y}|_{x=0,\,y=0} = d\mu_{SPDC}+d\mu_{SPDC}^2+d^2\left(\mu_{SPDC}+\mu_{PL}\right)^2,\\
    g^{(2)} = \frac{G^{(2)}_{12}}{ \langle N_{1}\rangle \langle N_{2}\rangle} = 1+\frac{\mu_{SPDC}^2}{d\left(\mu_{SPDC}+\mu_{PL}\right)^2}+\frac{d\mu_{SPDC}}{d^2(\mu_{SPDC}+\mu_{PL})^2} \approx 1+\frac{\alpha}{N_0}.
\end{gather}
Therefore, one can determine $\alpha$ by measuring the second-order normalized correlation function $g^{(2)}$ versus the detected number of photons in the first or the second arm of the HBT setup, $N^{det}_{1, 2} = \eta^{det}_{1, 2}N_0$, where $\eta^{det}_{1,2}$ stands for the detection efficiency of the corresponding arm:
\begin{equation}
    g^{(2)} = 1 + \frac{\eta^{det}_{1, 2}\alpha}{N^{det}_{1, 2}}.
\end{equation}
The value in the numerator is the heralding efficiency of arm $1,2$: $\eta_{1,2} = \eta^{det}_{1, 2}\alpha$.

In our experiment, through the time-resolved distillation of photon pairs we increase $\alpha$ from $0.085 \pm 0.002\%$ to $9.6\pm0.1\%$. 
Typical detected numbers of photons per pulse are $N^{det}_{1, 2}\approx 0.02$, corresponding to $N_0 \approx 0.2$ after accounting for the detection efficiency of $\eta^{det}_{1, 2}\approx 10\%$. Then, for this particular set of parameters, we show that the ultimate purity of the two-photon state that can be obtained with an ideal detection system is increased from 0.002 to 0.99.

\section{Efficiency budget}

In the experiment, through the time-resolved distillation of photon pairs from photoluminescence we increase the heralding efficiency $\eta_{1} = \eta_1^{det}\alpha$ by two orders of magnitude. Specifically, we increase $\alpha$ while the detection efficiency remains the same during the whole experiment. To estimate the absolute value of $\alpha$ directly from the heralding efficiency, one needs to know the detection efficiency $\eta_{1, 2}^{det}$. The latter, however, is quite uncertain in our experiment due to several aspects. Here, we estimate the detection efficiency of our setup by calculating the efficiency budget of our setup. 

The optical elements before the fiber detection system, which are lenses L1 and L2, two silver mirrors (not shown in the setup scheme), and spectral filters, account for about 10\% losses. The main source of losses in our setup is the following fiber-based detection system consisting of a single-mode 50:50 fiber beam splitter and two single-photon detectors coupled to single-mode fibers. Since SPDC radiation is highly multimode spatially~\cite{Okoth2019}, only a small fraction of all photon pairs emitted via SPDC in the collinear mode is transmitted through the fibers. Moreover, this mode must be accurately projected onto the mode of the fiber to reach the maximum coupling efficiency, which strongly depends on the alignment. Due to the complexity of coupling single photons to a single-mode fiber, the coupling efficiency is hard to estimate properly as it might vary a lot. Further, half of the photons is lost at the beamsplitter due to the events where both photons are directed into the same path. Finally, the detection efficiency of the SNSPDs is polarization dependent and varies from 50\% to 85\%, while the polarization of photons reaching the detectors is scrambled during the propagation through the fibers. Overall, the detection efficiency in our experiment has a high uncertainty: it is expected to be between 10\% to 30\% based on the loss estimation given above, which gives three times different estimations for $\alpha$. Therefore, we used a different approach to directly estimate the value of $\alpha$ by comparing the measurements for the different polarization configurations of emission.

\section{Efficiencies of different types of SPDC}
\begin{figure}[hbt!]
    \centering
    \includegraphics[width=0.8\linewidth]{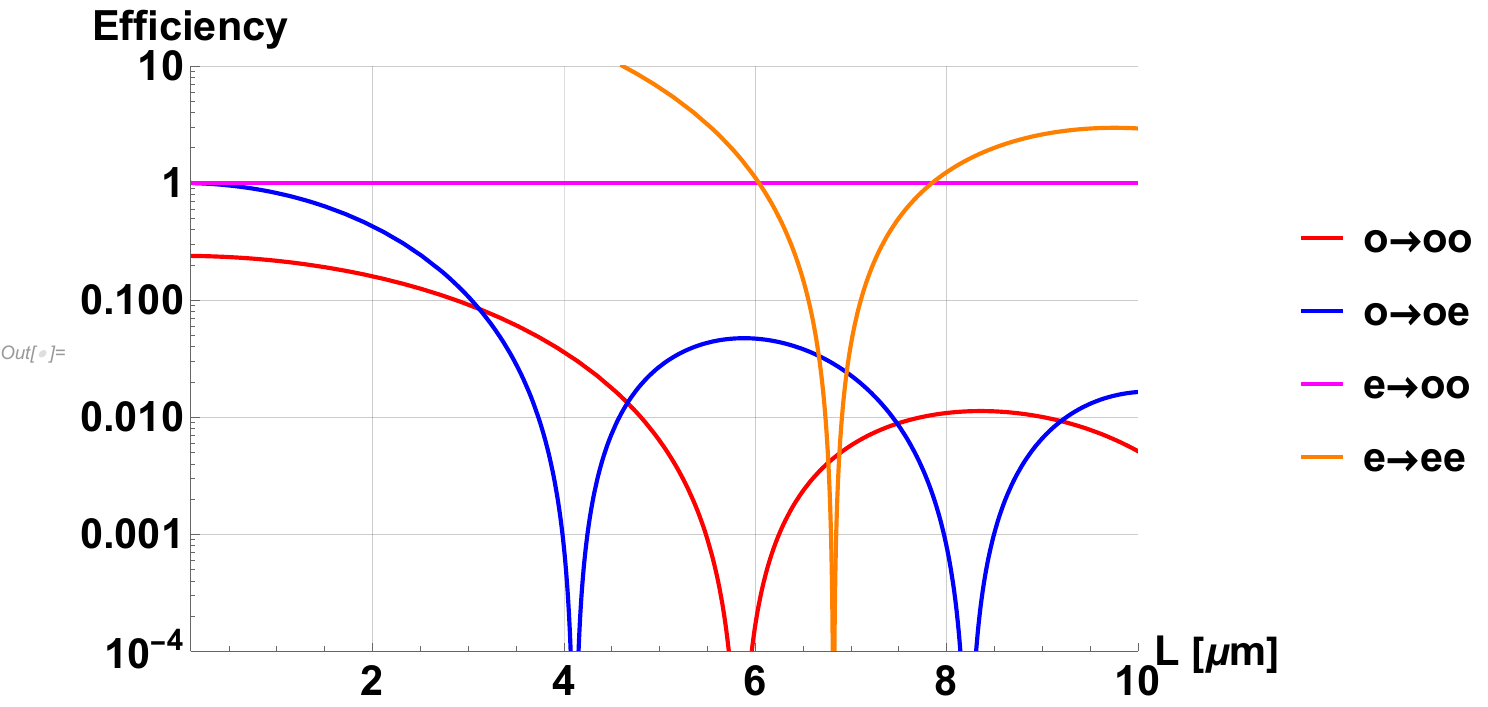}
    \caption{Photon pair generation efficiency for different polarization configurations strongly depends on the thickness of the LN wafer. All plots are normalized to the efficiency of the $e\rightarrow oo$ configuration, which is expected to be the highest. The thickness of the sample used in this work is about 7 $\mu$m.}
    \label{SPDC_efficiency}
\end{figure}

The proper way to check the ratio $\alpha$ between the rates of SPDC and the photoluminescence remaining after the distillation is to "switch off" SPDC and measure single counts and coincidences when only photoluminescence occurs. Photoluminescence is a linear process and generally depends on neither the pump polarization nor the polarization of the generated light. In contrast, the dependence of SPDC on the polarization configuration is strong, determined by the second-order nonlinear susceptibility tensor of the source and its thickness~\cite{Sultanov2022}. The second-order nonlinear tensor of LN used as a source of photon pairs in this work is~\cite{Boyd}
    \begin{equation}
        d = 
        \begin{pmatrix}
            0 & 0 & 0 & 0 & d_{31} & -d_{22} \\
            -d_{22} & d_{22} & 0 & d_{31} & 0 & 0 \\
            d_{31} & d_{31} & d_{33} & 0 & 0 & 0
        \end{pmatrix},
    \end{equation}
where $d_{22} \approx 2.1~\frac{pm}{V}$,  $d_{31} \approx -4.3~\frac{pm}{V}$, and $d_{33} \approx -34~\frac{pm}{V}$~\cite{Dmitriev}. In the X-cut LN thin film, there are four possible polarization configurations of SPDC listed in Table~\ref{LN_SPDC_polarizations}. Since LN is a uniaxial birefringent crystal with the optic axis along the z direction, let us call z-polarized photons extraordinary (e-) polarized and y-polarized photons ordinary (o-) polarized.
\begin{table}[hbt!]
    \centering
    \begin{tabularx}{1\textwidth}{ | >{\centering\arraybackslash}X | >{\centering\arraybackslash}X | >{\centering\arraybackslash}X | >{\centering\arraybackslash}X |}
    \hline
    Polarization configuration & Related component of the nonlinear tensor & SPDC phase-matching type & $L_{coh}$, $\mu m$\\
    \hline
    $o\longrightarrow o~o$ & $d_{22}$ & type-0 & 2.92\\
    \hline
    $o\longrightarrow o~e$ & $d_{31}$ & type-II & 2.05\\
    \hline
    $e\longrightarrow o~o$ & $d_{31}$ & type-I & 185\\
    \hline
    $e\longrightarrow e~e$ & $d_{33}$ & type-0 & 3.41\\
    \hline
    \end{tabularx}
    \caption{Polarization configuration of SPDC in X-cut LN.}
    \label{LN_SPDC_polarizations}
\end{table}
For conventional SPDC in bulk crystals, usually, only one type of SPDC occurs at a time due to phase matching. In the case of non-phase-matched SPDC, all four configurations are possible, and their relative efficiencies depend on the film thickness and the value of the related nonlinear tensor component, even though the thickness of the crystal used in this work is on the order of several coherence lengths for three out of four polarization configurations. To estimate the efficiency of SPDC in different polarization configurations, it is enough to plot the efficiency for every polarization configuration as a function of effective nonlinear tensor $\chi_{eff}$, crystal thickness $L$, and longitudinal phase mismatch $\Delta k$,
\begin{equation}
    \epsilon = \chi^2_{eff}L^2sinc^2\left(\frac{\pi L}{2 L_{coh}}\right),
\end{equation}
where $L_{coh}=\pi/\Delta k$ is the coherence length.
The efficiencies for all four polarization configurations are shown in Fig.~\ref{SPDC_efficiency} as functions of the LN thickness $L$. Type-I SPDC with configuration $e\rightarrow oo$ is expected to be the strongest, although the corresponding effective nonlinear tensor is one order of magnitude smaller than that for the $e\rightarrow ee$ configuration. This is a consequence of the crystal length being close to two coherence lengths for the $e\rightarrow ee$ configuration, which is significantly suppressed in this case.
As we can see from Fig.~\ref{SPDC_efficiency}, for the length of of the crystal of 7 $\mu$m, the SPDC efficiency for the o-polarized pump is negligible compared to the strongest interaction ($e\rightarrow oo$). Therefore, we assume in the main manuscript that all single counts and coincidences obtained with this pump polarization are due to photoluminescent photons. This upper-boundary estimation for the residual photoluminescence already demonstrates a high efficiency of the time-resolved distillation of photon pairs. In the main manuscript, we estimate $\alpha$ for the most efficient process ($e\rightarrow oo$) to be $90\%$ when the time-resolved distillation is applied, by comparing the three-fold coincidence rates for $e\rightarrow oo$ and $o\rightarrow ee$. Therefore, the relatively low value of the heralding efficiency, about 10\%, is entirely determined by the detection efficiency, so that $\eta_{1}^{det}$ is estimated to be 10\%. Without the time-resolved distillation, the fraction $\alpha$ of photons generated via SPDC does not exceed 1\%.

\section{Two-photon spectrum}

\begin{figure}[hbt!]
    \centering
    \includegraphics[width=0.8\linewidth]{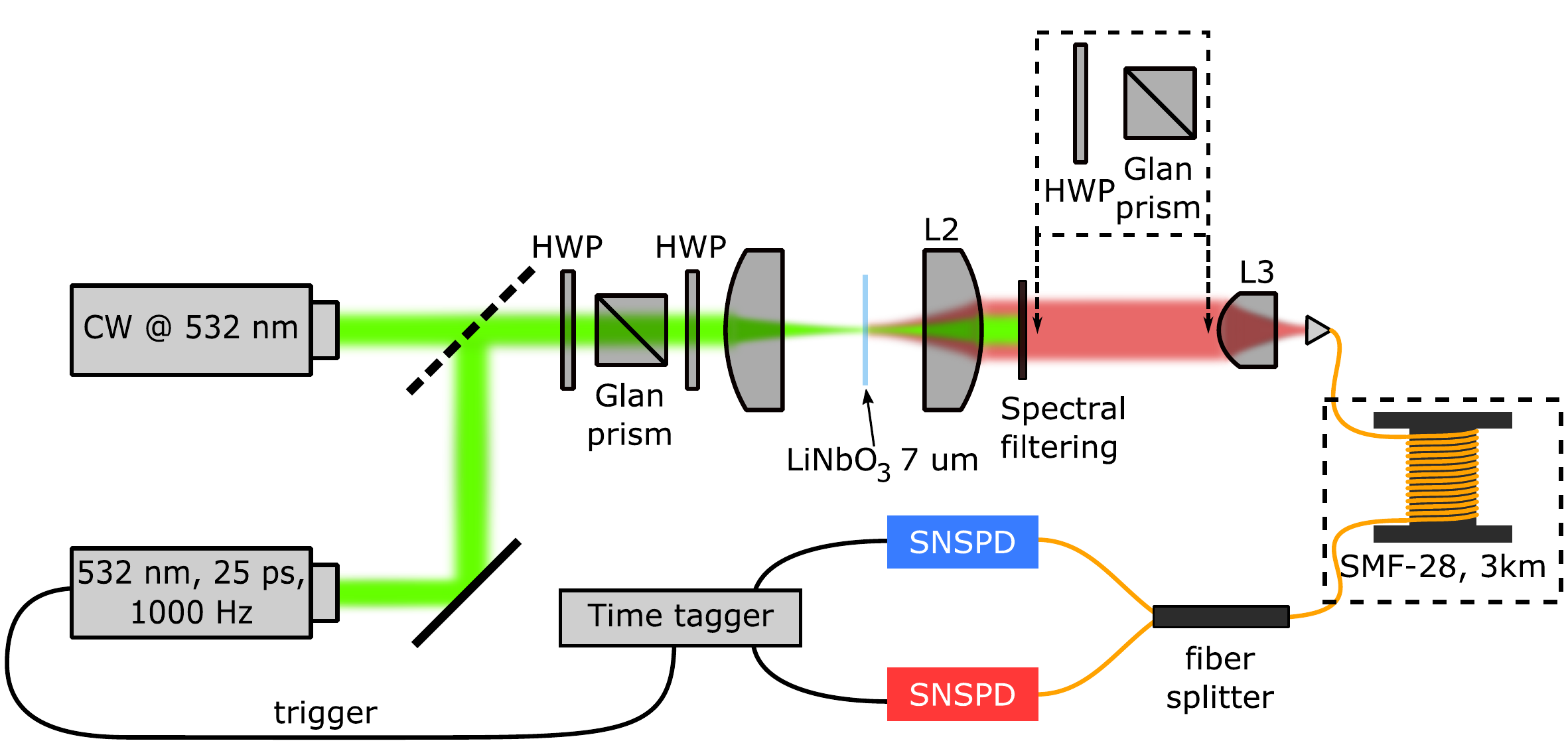}
    \caption{For the two-photon fiber spectroscopy, we place 3 km of SMF-28 before the beamsplitter and disperse the two-photon wave packet, mapping the SPDC spectrum to the coincidence histogram.}
    \label{Setup_FS}
\end{figure}

\begin{figure}[hbt!]
    \centering
    \begin{subfigure}{0.45\linewidth}
        \includegraphics[width=1\linewidth, valign=b]{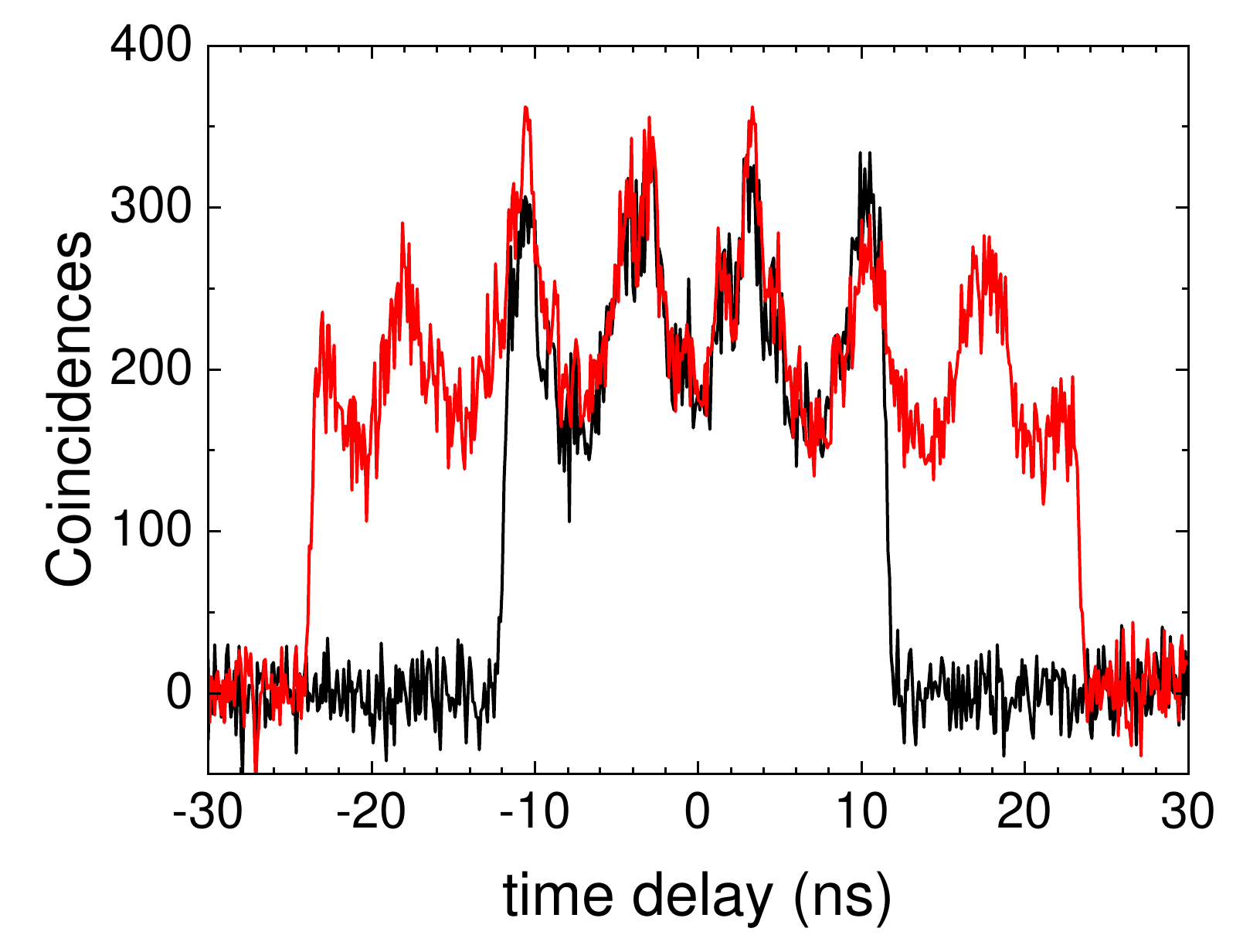}
        \caption{}
        \label{Uncalibratedspectra}
    \end{subfigure}
    \hspace{1cm}
    \begin{subfigure}{0.45\linewidth}
        \includegraphics[width=1\linewidth, valign=b]{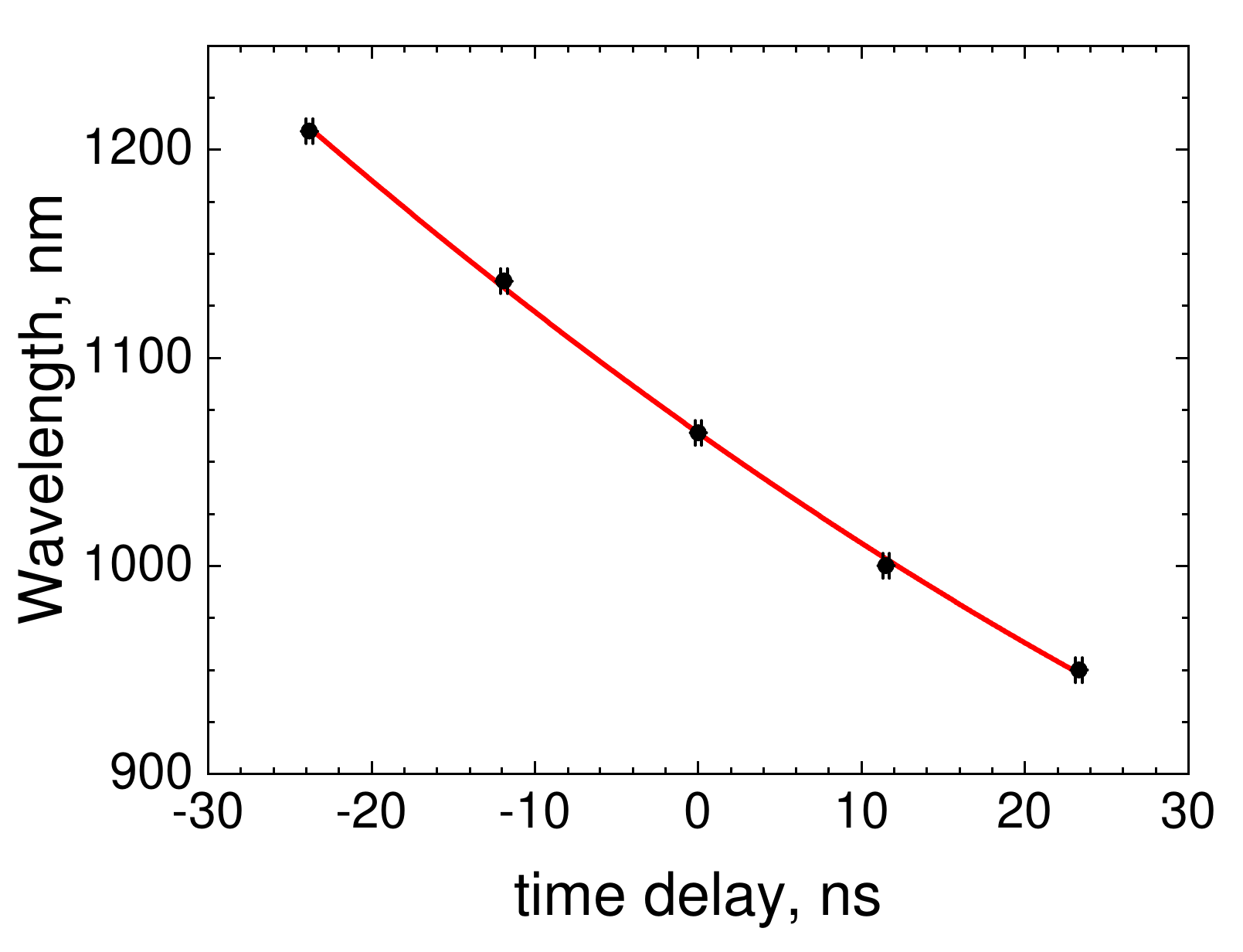}
        \caption{}
        \label{Calibration_curve}
    \end{subfigure}
    \caption{We measure the coincidence histogram with two different long-pass filters, with the cut-on wavelength 1000 nm (black in Fig.~\subref{Uncalibratedspectra}) and 950 nm (red in Fig.~\subref{Uncalibratedspectra}). The background is subtracted. Then, we relate the time delays of the spectra edges to the cut-on wavelengths of the filter and the conjugated wavelengths and obtain a calibration curve (Fig.~\subref{Calibration_curve}).}
    \label{Calibration}
\end{figure}

Finally, we measured the two-photon spectrum to estimate the number of frequency modes. We use single-photon fiber spectroscopy to resolve the two-photon spectrum through the correlation measurements~\cite{Valencia2002}. As a dispersive medium, we use a 3 km long SMF-28 fiber (Fig.~\ref{Setup_FS}). The fiber spreads a two-photon wave packet, mapping the SPDC spectrum to the histogram of time delays between the arrival times of two photons (the coincidence histogram).
We measure the coincidence histogram with two different long-pass filters LPs (Fig.~\ref{Uncalibratedspectra}) and map the time delays of the obtained histograms to the edges of the expected spectra, which are defined by the cut-on wavelengths of the LPs. While the cut-on wavelength of a filter corresponds to one edge of the spectrum, the second edge corresponds to the conjugated wavelength defined by the energy conservation in SPDC. Since both photons travel through the same fiber and experience identical dispersion, the time arrival difference for the degenerate case remains unchanged. Therefore, we also use the fifth point that corresponds to the frequency-degenerate case and the time delay between photons detected without the fiber. Then, we fit the obtained data by a second-order polynomial and obtain a calibration curve to map the wavelengths onto the time delays (Fig.~\ref{Calibration_curve}).

\begin{figure}[hbt!]
    \centering
    \includegraphics[width=0.8\linewidth]{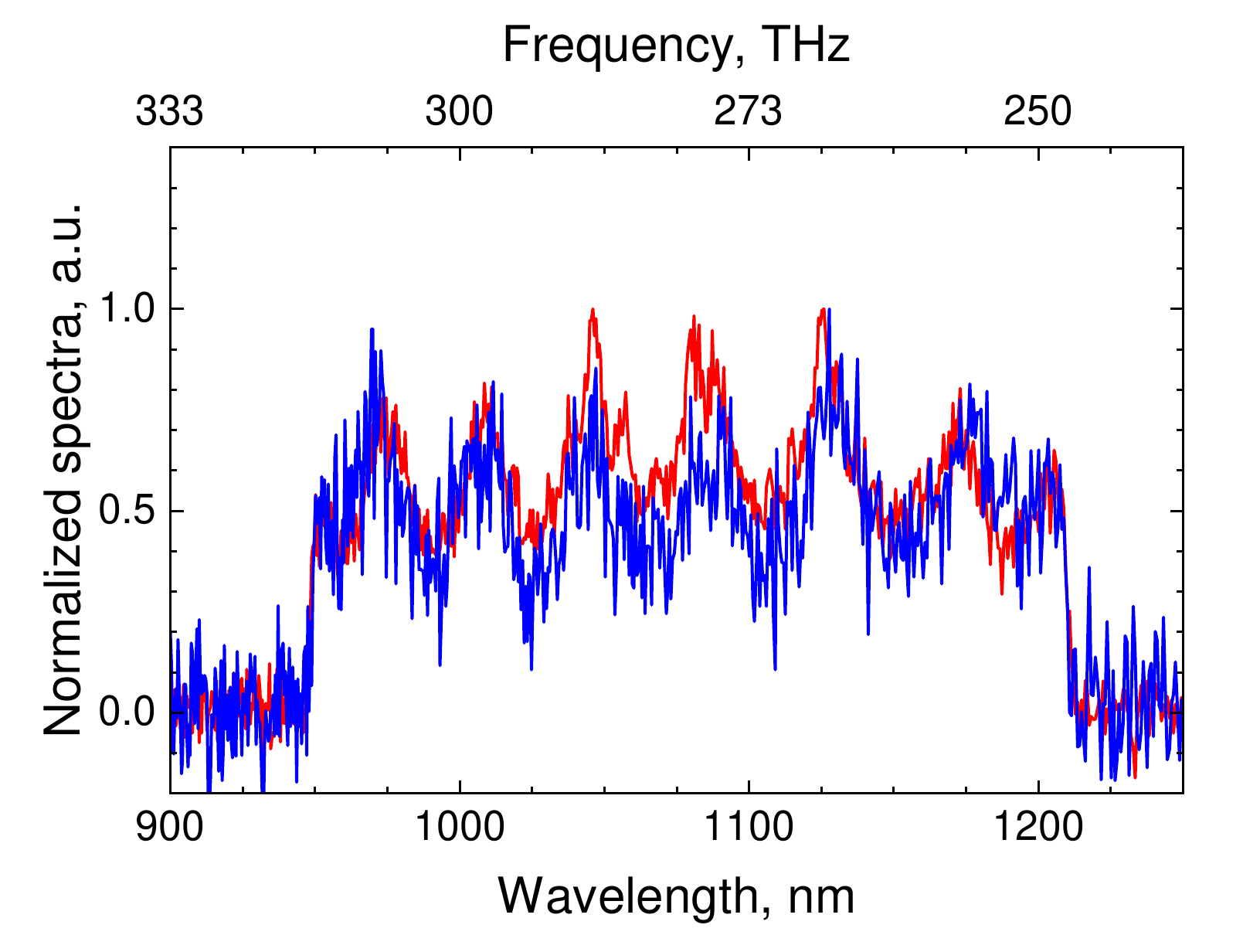}
    \caption{SPDC spectrum measured for the $e\rightarrow ee$ (blue curve) and $e\rightarrow oo$ (red curve) configurations.}
    \label{Spectra}
\end{figure}

The measured spectra are shown in Fig.~\ref{Spectra}. For both polarization configurations, the spectra look identical, and their modulation is caused by the etalon effect in the source~\cite{Kitaeva2004}. Both the relative efficiencies of these two processes and the two-photon spectra can be adjusted by changing the thickness of the source to achieve polarization entanglement within a broad frequency range and therefore generate a hyper-entangled state.
The measured SPDC spectra stretch from 950 nm to  1210 nm, covering 68 THz. The number of frequency modes can be roughly estimated as the Fedorov ratio, which is the ratio between the spectral width $\Delta\nu$ of SPDC and the correlation width $\delta\nu$, given by the spectral width of the pump~\cite{Fedorov}:
\begin{equation}
    R = \frac{\Delta\nu}{\delta\nu}.
\end{equation}
By taking $\delta\nu$ equal to the linewidth of the pulsed pump in our experiments (0.06 THz), and $\Delta\nu = 68$ THz, we obtain $R=1130$, which is also the number of modes $d$ that we use to estimate the purity of the state.